\documentclass[12pt, a4paper]{article}
\usepackage{amsfonts}
\usepackage{amsmath}
\usepackage{epsfig}
\usepackage{latexsym}
\usepackage{amssymb}
\usepackage{natbib}
\usepackage[nodots]{numcompress}
\usepackage[version=3]{mhchem}
\usepackage[usenames, dvipsnames]{xcolor}
\usepackage{soul}

\def\b1{{\mathbf 1}}

\begin{document}

\title{River flow modelling using nonparametric functional data analysis 
\thanks{%
{This is the peer reviewed version of the following article: [Quintela-del-R\'io, A., Francisco-Fern\'andez, M. (2018). River flow modelling using nonparametric functional data analysis. Journal of Flood Risk Management, 11(S2), S902-S915. DOI: 10.1111/jfr3.12282], which has been published in final form at: https://doi.org/10.1111/jfr3.12282. This article may be used for non-commercial purposes in accordance with Wiley Terms and Conditions for Use of Self-Archived Versions. This article may not be enhanced, enriched or otherwise transformed into a derivative work, without express permission from Wiley or by statutory rights under applicable legislation. Copyright notices must not be removed, obscured or modified. The article must be linked to Wiley's version of record on Wiley Online Library and any embedding, framing or otherwise making available the article or pages thereof by third parties from platforms, services and websites other than Wiley Online Library must be prohibited.}}}
\author{Alejandro Quintela-del-R\'{\i}o \\
University of A Coru\~na \and Mario Francisco-Fern\'andez \\
University of A Coru\~na \thanks{%
Faculty of Computer Science, Campus de Evi\~na, s/n, A Coru\~na 15071,
Spain. Email: mariofr@udc.es.} }

\date{}
\maketitle

\begin{abstract}
Time series and extreme value analyses are two statistical
approaches usually applied to study hydrological data. Classical techniques,
such as $ARIMA$ models (in the case of mean flow predictions), and parametric
generalised extreme value (GEV) fits and nonparametric extreme value methods
(in the case of extreme value theory) have been usually employed in this
context. In this paper, nonparametric functional data methods are used to
perform mean monthly flow predictions and extreme value analysis, which are important for flood risk management. These are
powerful tools that take advantage of both, the functional nature of the
data under consideration and the flexibility of nonparametric methods,
providing more reliable results.
Therefore, they can be useful to prevent damage caused by floods and to reduce the likelihood and/or the impact of floods in a specific location.
 The nonparametric functional approaches are
applied to flow samples of two rivers in the U.S. In this way, monthly mean flow is predicted
and flow quantiles in the extreme value framework are estimated using the
proposed methods. Results show that the nonparametric functional techniques work
satisfactorily, generally outperforming the behaviour of classical parametric
and nonparametric estimators in both settings. \newline
\noindent \emph{Key Words}: Nonparametric estimation, Functional
data, River flow, Forecasting, Extreme values
\end{abstract}

\renewcommand{\baselinestretch}{1.5}

\section{Introduction}

\label{intro}

Prediction of future values
is essential for the design of water systems, and control
measures will be more effective if the process is reliable.
Likewise, management and scheduling of areas exposed to
flood risk rely heavily on tools for frequency analysis of hydrological
extremes.

Numerous studies have been carried out on hydrological problems using
statistical methods. Among them, time series prediction is
topical in this field \citep{Toth00,Tamea05,Wu09}. 
Research studies on time series also include linear models for forecasting
river flows 
\citep[see,
e.g.,][and references therein]{Wang09}.  
Among the several techniques to model time series, autoregressive integrated moving average ($ARIMA$) models described well the data analysed in the present research and, therefore, they were employed to fit the hydrological time series studied. Moreover, with this choice, a similar comparison (between $ARIMA$ models and nonparametric functional methods) to that performed in some works widely cited in the literature \citep{sankhya2005} can be carried out.
Basically, $ARIMA$ models are preferred for time series of short-memory type
(the autocorrelation structure decreases quickly), while, in
other cases, hydrological processes are of long-memory type.
Other possible alternatives not considered in the present research would be, for
example, fractional Gaussian noise and broken line models \citep{Kout00}.

Statistics of extremes \citep{Coles01}
is also one of the most significant techniques in frequency analysis 
\citep[see,
e.g.,][]{Kat02, Saf09, Singh05}. Daily, monthly or annual maximum time
series of river flow recordings are typically represented by the generalised
extreme value (GEV) distribution.

$ARIMA$ and GEV fitting are typical examples of parametric modelling. A different type of statistical model
applied to hydrological data involves using nonparametric curve estimation
methods, which does not require
restrictive assumptions on the distribution of the population of
interest. Several papers have applied nonparametric estimation methods to
hydrological time series to carry out predictions as well as to perform extreme value analysis
\citep{Lall93, Guo96, Sharma97,
Kim02, Wang09, Quintela2011}. Further details regarding
nonparametric techniques (including theoretical motivations, practical
applications to several scientific fields and references) may be found in,
for instance, the books of \cite{Ruppert2003} or \cite{Wasserman05}.

Time series were recently analysed by nonparametric functional data
analysis (NFDA) \citep{ferratyvieulibro, ramsaysilvermanlibro}. 
NFDA works with data consisting of curves or multidimensional variables.
Different procedures using these techniques have
been applied to several complex real problems %
\citep{besse2000,hallposkitt2001,fernandezdecastro2005, castellano2009}.

The present paper focuses on applying NFDA techniques in prediction
problems and extreme value analysis in the setting of hydrology. 
The organisation of the paper is as follows. Section~\ref{stat-methods} presents the statistical methods used in this
paper.
These methods correspond to $ARIMA$ 
models for time
series prediction (Section \mbox{\ref{pred-mod-arima}}), and GEV parametric
estimators and nonparametric methods in extreme value analysis (Sections \mbox{\ref{ext-val-an}} and \mbox{\ref{nopara}}, respectively).
Next, the new
proposals using NFDA to study both problems (time series prediction and
extreme value analysis) are presented (Section \ref{pred-app}).
In Section~\ref{application}, these
tools are applied to river flow data from two sites in the US. Finally, in Section \ref{conclussions}, a general discussion of the results is included.

\section{Statistical methods}

\label{stat-methods}

\subsection{Time series analysis. $ARIMA$ models}

\label{pred-mod-arima}

Let $\{Z_{t}\}_{t\in 
\mathbb{R}}$ be a stochastic process or time series observed until a time $T$.
Usually, the process is observed at $N$ discretised times, and the
observations are denoted by $\{Z_{1}, \ldots ,Z_{N}\}$. To predict a future value 
$Z_{N+s}$, the simplest way consists of taking into account one single past
value. This is done by constructing a two-dimensional statistical sample of
size $n=N-s$, by setting $X_{i}=Z_{i}$ and $Y_{i}=Z_{i+s}$, with $%
i=1,...,N-s $. Therefore, the problem is converted into a standard
prediction problem of a response $Y$, given an explanatory variable $X$.
This can be generalised by considering the following autoregressive process
of order $p$: 
\begin{equation}
Z_{i+s}=m(Z_{i},\dots ,Z_{i-p+1})+\varepsilon _{i},\text{ }i=p, \ldots ,N-s,
\label{sample2}
\end{equation}%
where $\varepsilon _{i}$ is the error process, assumed to be independent of $%
Z_{i}$, and the aim is to estimate the function $m(\cdot )$.

A first approximation consists in assuming that $m(\cdot )$ belongs to a
particular class of functions, only depending on a finite number of
parameters to be estimated, such as the $ARIMA(p,d,q)$ models \citep{Singh05}.
If $d$ is a non-negative integer, then $\{Z_t\}$ is said an $ARIMA(p,d,q)$ process 
if $Y_t=(1-B)^{d}Z_t$ is a causal $ARMA(p,q)$ process, where $B$ is the backward shift operator defined by $B^jZ_t=Z_{t-j}$, $j=0,\pm 1,\pm 2, \ldots$
\citep{Brockwell91}.
Note that the process $\{Z_t,t=0, \pm 1 \pm 2, \ldots\}$ is said an $ARMA(p,q)$ process if $\{Z_t\}$ is stationary and if for every $t$,
$$
Z_t-\phi_1 Z_{t-1} - \ldots - \phi_p Z_{t-p}=\varepsilon_t + \theta_1 \varepsilon_{t-1} + \ldots + \theta_q \varepsilon_{t-q},
$$
where $\{\varepsilon_t \}$ is a process of error terms, generally assumed to be uncorrelated random variables with mean 0 and variance $\sigma^2$.
In a $ARIMA(p,d,q)$, $p$ is the order (number of time lags) of the autoregressive model, $d$ is the degree of differencing (the number of times the data have had past values subtracted), and $q$ is the order of the moving-average model.
The good practical properties of $ARIMA$ models have led to regularly use
them to study hydrological problems. Some relevant papers on this topic are,
for example, \cite{Montanari97}, \cite{Toth00} or \cite{Tamea05}.

The
prediction problem can be tackled using nonparametric methods.
To apply these methods only some mild regularity conditions on the function $m(\cdot )$ have to be assumed.
The ``curse of dimensionality problem" \citep[][p. 90]{wan95} is particularly troublesome in this nonparametric framework.
It has to do with the selection of the number of past values to consider in
the model. This is indeed an important question. The lower the number of
past predictors, the less flexible the model is, but when the lag 
increases, a large number of observations are needed to obtain good estimates
of the model parameters. This number increases exponentially as the
dimension becomes larger.

\subsection{Extreme value analysis}

\label{ext-val-an_both}

\subsubsection{The GEV distribution}

\label{ext-val-an}

Suppose $X_{1}, \ldots ,X_{n}$ is a sequence of extreme values with a common
distribution function $F$. In the context of the present paper, these
variables can be the maximum river flows measured in a specific period
of time (24 h, a month, a year, etc.). Classical parametric extreme value
theory uses the idea that, under certain regularity conditions %
\citep{Fisher28}, the limit of the distribution function $F$ of the maximum
is the GEV distribution. 
Its cumulative distribution function is:
\begin{equation}
F_{\theta}(x)=\left\{ 
\begin{tabular}{ll}
$\exp \left\{ -[1+\gamma (x-\mu )/\sigma ]^{-1/\gamma }\right\} $ & if $\gamma \neq 0$\\  
$\exp \{-\exp [-(x-\mu )/\sigma ]\}$ & if $\gamma =0$%
\end{tabular}%
\right.  \label{gev}
\end{equation}
with $\theta =(\mu ,\sigma ,\gamma )$. Here, $\mu $ is the location
parameter, $\sigma >0$ is the scale parameter, and $\gamma $ is the shape
parameter. Mean and standard deviation are obtained as functions of these
parameters \citep[see, e.g.][]{Coles01}. 
The range of definition of the GEV distribution depends on $\gamma$. If $\gamma \neq 0$, $F_{\theta}(x)$ is defined for $x$ such that $1+\gamma (x-\mu )/\sigma >0$, while if $\gamma =0$, it is defined for $-\infty <x< \infty$.
Various values of the shape parameter yield the extreme value type I, II, and III distributions. Specifically, the three cases $\gamma=0$, $\gamma>0$, and $\gamma<0$ correspond to the Gumbel, Fr\'echet, and ``reversed" Weibull distributions, respectively.
Using the random sample of extreme
values, an estimator $\hat{\theta}$ for $\theta $ can be obtained. 
Then, substituting $F$ by $F_{\hat{\theta}}$, estimators of some important
functions in this framework can be defined. For instance:

\begin{itemize}
\item The function providing the probabilities of exceedance. In the context
of the present paper, it corresponds to the function that, for a river flow $%
c$, gives the probability of obtaining a flow larger than $c$ (per unit of
time). It is defined as 
\begin{equation}
R(c)=P(X>c)=1-F(c).  \label{risk}
\end{equation}

\item The flow quantile, defined as
the value of the flow that can be expected to be once exceeded during a $T$%
-period of time. For each value of $T$, it is given by 
\begin{equation}
FQ(T)=F^{-1}\left( 1-\frac{1}{T}\right)  \label{rl}
\end{equation}

\item The mean return period or recurrence interval of a particular river flow
$c$, defined as an estimator of the interval of time between events of
this flow. It can be expressed as the inverse of the probability that a
flow $c$ will be exceeded in one period of time: 
\begin{equation}
RT(c)=\frac{1}{P(X>c)}=\frac{1}{1-F(c)}.  \label{mrp}
\end{equation}

An application of these expressions is given in Section \ref{secondcase}.
\end{itemize}

\subsubsection{Nonparametric estimators}

\label{nopara} The main advantage of working with nonparametric methods is
that they are model-free, that is, no specific
functional form is required for the parameters or curves to be estimated. Several
nonparametric estimators for different functions of interest have been
developed in the last decades. In this work, kernel estimators of the
density function and the distribution function are used.

Let $X$ be a continuous random variable, with density function $f$ and
distribution function $F.$ Given a random sample $X_{1}, \ldots ,X_{n},$ each $%
X_{i}$ having the same distribution as $X$, the Parzen-Rosenblatt
nonparametric kernel estimator \citep{Parzen62} of $f$ is defined by: 
\begin{equation}
f_{h}(x)=\dfrac{1}{nh}\sum_{i=1}^{n}K\left( \dfrac{x-X_{i}}{h}\right),
\label{kernel}
\end{equation}%
where $K$ is a kernel function (normally, $K$ is a density function with
some regularity conditions) and $h=h(n)\in \mathbb{R}^{+}$ is the smoothing
parameter (or bandwidth) that regulates the amount of smoothing to be used.
From the relation between a density function and a distribution function, a nonparametric kernel estimator of the
distribution function can be directly constructed: 
\begin{equation}
F_{h}(x)=\int_{-\infty }^{x}f_{h}(t)dt=\dfrac{1}{n}\sum_{i=1}^{n}H\left( 
\dfrac{x-X_{i}}{h}\right) ,  \label{kerneldist}
\end{equation}%
where $H(u)=\int_{-\infty }^{u}K(t)dt$ is the distribution function of the
kernel $K$.

Using equation (\ref{kerneldist}),
nonparametric estimators of the probabilities of exceedance, the flow
quantiles, and the recurrence intervals defined in (\ref{risk}), (\ref{rl})
and (\ref{mrp}), respectively, can be obtained: 
\begin{equation}
R_{h}(c)=1-F_{h}(c),  \label{risknopara}
\end{equation}%
\begin{equation}
FQ_{h}(T)=F_{h}^{-1}\left( 1-\frac{1}{T}\right) .  \label{rlnopara}
\end{equation}%
and 
\begin{equation}
RT_{h}(c)=\frac{1}{1-F_{h}(c)}.  \label{mrpnopara}
\end{equation}

An important first step to compute (\ref{risknopara}), (\ref{rlnopara}) and (%
\ref{mrpnopara}) is the selection of the smoothing parameter $h$. Popular
techniques to select the bandwidth are the modified cross-validation %
\citep{Bowman98, Quintela2011} and plug-in methods 
\citep{Lall93,
Quintela2011}. In the examples presented in this work, a
cross-validation bandwidth selection method is used.

In an extreme value framework, it can be of interest to estimate the %
flow quantiles or the return periods for
extremely large events. In a hydrological context, \cite{Lall93} found that the previous
nonparametric estimators can suffer from boundary problems. Some authors
have addressed extrapolation issues using nonparametric estimators like
those given in (\ref{rlnopara}) or (\ref{mrpnopara}). They basically focused
on studying the influence of the kernel function and the bandwidth parameter
in the final results. Regarding the kernel, while \cite{Guo96} proposed to
use a Gumbel kernel and \cite{Lall93} discussed the use of a variable kernel
distribution function estimator to tackle this problem, \cite%
{AdamowskiFeluch90} tested Gaussian, Gumbel and Epanechnikov kernels in
flood frequency analysis and found that the choice of the kernel is not
important, and the shape of the kernel does not affect extrapolation
accuracy. As for the smoothing parameter, the use of variable or local
bandwidths to address the extrapolation problem was discussed in \cite%
{Adamowski89} or \cite{Guo96}. 
Note that the variance of the (parametric or nonparametric)
estimators can increase significantly when the interest is to estimate extremely large flow quantiles. For this, in that case,  the results obtained should be considered carefully. In the present
paper, the methods will always be applied for values inside
the range of the observed data.

Nonparametric kernel quantile function estimators based on smoothing the
empirical quantile function are proposed and studied by \cite{MoonLall94}
and \cite{apipa}. They follow similar ideas, but while in \cite{MoonLall94}
the Gasser--M\"uller estimator \citep{gasser} with higher order kernel is
used, in \cite{apipa} the smoothing process is carried out employing the
local polynomial estimator \citep{fan96} with a local bandwidth.

\subsection{Functional data. NFDA techniques}

\label{fun-data-tec}Let $\{(\chi _{i},Y_{i}),\,i=1,\ldots, n\}$ be a sample
of $n$ random pairs, each distributed as $(\mathcal{X},Y)$, where the
variable $\mathcal{X}$ is of functional nature (a curve), and $Y$ is scalar.
Formally, $\mathcal{X}$ is a random variable valued in some
semi-metric functional space $E$, and $d(\boldsymbol{\cdot},%
\boldsymbol{\cdot})$ denotes the associated semi-metric, according to the definition \citep{ferratyvieulibro}:

\begin{enumerate}
\item $\forall \mathbf{x}\in E,\ d(\mathbf{x},\mathbf{x})=0.$

\item $\forall \mathbf{x},\mathbf{y},\mathbf{z}\in E,\ d(\mathbf{x},\mathbf{y%
})\leq d(\mathbf{x},\mathbf{z})+d(\mathbf{z},\mathbf{y}).$
\end{enumerate}

The conditional cumulative distribution of $Y$ given $\mathcal{X}$ is
defined for any $y\in
\mathbb{R}$ and any $\mathbf{\chi }\in E$ by:%
\begin{equation}
F(y|\mathbf{\chi })=P(Y\leq y|\mathcal{X}=\mathbf{\chi }).  \label{fdist}
\end{equation}

A functional variable can be considered a generalisation of a
mul\-ti\-di\-men\-sio\-nal variable, assuming that the variable $\chi $ is $%
p $-dimensional, with $p$ an integer (for example, $p=12$ for the monthly
mean flow in the twelve months of a year). In this case, the functional
space would be $E=
\mathbb{R}^{p}$ and the semi-metric could be the classical Euclidean distance or some
equivalent measure \citep{ramsaysilvermanlibro}.

Both parametric and nonparametric methods can be used in functional data
applications. The monograph of \cite{ferratyvieulibro} provides a benchmark of
nonparametric curve estimation for functional data. As shown in this book,
the conditional distribution $F(\cdot |\mathbf{\chi })$ given in (\ref{fdist}%
) can be nonparametrically estimated by:%
\begin{equation}
\hat{F}_{n}(y|\mathbf{\chi })=\frac{\sum_{i=1}^{n}K\left( \frac{d(\chi ,\chi
_{i})}{h}\right) H\left( \frac{y-Y_{i}}{g}\right) }{\sum_{i=1}^{n}K\left( 
\frac{d(\chi ,\chi _{i})}{h}\right) },  \label{est-dist}
\end{equation}%
where $K$ is a kernel function and $H$ is defined as the distribution of
another kernel density function $K_{0},$ that is, $H(x)=\int_{-\infty
}^{x}K_{0}(u)du$. Parameters $g$ and $h$ are smoothing parameters or
bandwidths (they could take the same value).

Expression (\ref{est-dist}) is a direct extension of the nonparametric
estimator of a conditional distribution function ($F(y|X=x)$, for $(X,Y)$
real random variables \citep{hallwolffyao}). 
The main difference
between functional and non-functional estimators lies in the use of a
semi-metric $d(\chi ,\chi _{i})$ instead of the Euclidean distance $%
\left\Vert \chi -\chi _{i}\right\Vert $. Several types of kernel functions
and semi-metrics can be considered 
\citep[see][Sections
3.2-3.4]{ferratyvieulibro}, depending, essentially, on the data at hand.
Theoretical optimality properties of the estimator (\ref{est-dist}) can be
found in \cite{quintela2008}.

An important advantage of NFDA techniques is that the framework model
reduces to a bivariate setting and, therefore, the curse of dimensionality
problem is basically avoided. Additionally, the boundary problems of
nonparametric estimators, previously described, can be partially avoided in
functional data estimation. This fact requires a proper choice of the
semi-metric \citep{ferratyvieulibro}.

\subsubsection{NFDA applied to time series analysis}

\label{pred-app} As it is well known, $ARIMA$ models are constrained by their
particular structure and the number
of past values used in the statistical model for prediction purposes. NFDA
methods overcome these two restrictions, because of the nonparametric nature
of the approaches and dividing the observed seasonal time series into a
sample of curves. 
In Section \ref{firstcase}, NFDA methods are applied to predict monthly mean flows in practical situations, and the performance of these approaches is compared with that obtained when $ARIMA$ models are employed.

To analyse a monthly mean series as a set of functional data, the original
time series is converted into annual curves. Note that if there were some
months in which the corresponding measures were not available, the curves
would not have the same number of components (this is known as an unbalanced
data setting), and more complex specific preprocessing would be required 
\citep[see
Section 3.6 of][]{ferratyvieulibro}. 
Let $%
\{Z_{k}\}_{k=1}^{N}$ be the complete time series. For $%
i=1,\ldots ,n$, the annual curves, $\chi _{i}$ $=(\chi _{i}(1),\ldots ,\chi
_{i}(12))$, are constructed, where 
\begin{equation}
\forall t\in \{1,2,\ldots ,12\},\ \ \ \ \ \chi _{i}(t)=Z_{12\cdot (i-1)+t},
\label{f1}
\end{equation}%
corresponds to the monthly mean flows of the $i$th year. 
Each annual curve is considered as a continuous path (i.e. $\chi _{i}=$ $\{Z_{12\cdot
(i-1)+t},t\in \lbrack 0;12)\}),$ but observed only at 12 discretised points.
Thus, the time series consists of a sample of $n$ dependent functional data $%
\chi _{1},\ldots ,\chi _{n}$.

In this way, much information from the past
of the time series can be taken into account, but still using for the past a single continuous object
(exactly one year). For more insight on this issue, let us suppose, for
instance, that the time series could be measured $p$-times each year with $%
p>12$. In this case, the functional data
analysis will consider the whole continuous past year and the same
asymptotic behaviour remains, independent of $p$. 

In order to predict the monthly mean flow in the year $n+1$, the following process was carried out. For $i=1,\ldots ,n-1$ and for any fixed $\delta $ in $%
\{1,\ldots ,12\}$, take $Y_{i}(\delta )=\chi _{i+1}(\delta )$, i.e., $%
Y_{i}(\delta )$ denotes the monthly flow in the month $\delta $ and the year 
$i+1$. Thus, a sample of $n-1$ pairs, $\{\chi _{i},Y_{i}(\delta
)\}_{i=1}^{n-1}$, with $Y_{i}(\delta )$ a real variable and $\chi _{i}$ a
functional one, is available. According to Section \ref{fun-data-tec}, a predictor of $%
Y_{n}(\delta )$, knowing $\chi _{n}$, can be achieved by estimating the
median of the conditional distribution:%
\begin{equation}
\hat{Y}_{n}(\delta )=\hat{t}_{0.5}=\hat{F}_{n}^{-1}(0.5 | \chi _{n}),
\label{med1}
\end{equation}%
where $\hat{F}_{n}(\cdot | \chi _{n})$ is the estimated distribution of $%
Y(\delta )$ given $\chi _{n}$. Repeating this step for $\delta =1,\ldots
,12,$ the mean values of the flow for the $(n+1)$th year can be predicted. 

In the functional data
context of this paper, another approximation consists in considering a regression model like (\ref{sample2}) and using a nonparametric kernel  functional method to estimate the regression function $%
m(\cdot )$.
Considering the sample data of functional covariates and
a scalar response, $\{\chi _{i},Y_{i}(\delta
)\}_{i=1}^{n-1}$, the nonparametric functional estimator %
\citep{ferratyvieulibro} has the expression: 
\begin{equation}
\hat{m}(\chi )=\frac{\sum_{i=1}^{n-1}Y_{i}(\delta )K\left( \frac{d(\chi ,\chi
_{i})}{h}\right) }{\sum_{i=1}^{n-1}K\left( \frac{d(\chi ,\chi _{i})}{h}\right) 
}.  \label{noparareg}
\end{equation}

Equation (\ref{noparareg}) constitutes a functional alternative based on
regression techniques to the approach previously used based on median
estimation. Using (\ref{noparareg}), the flows of the $(n+1)$th year can be
predicted calculating 
\begin{equation}
\hat{Y}_{n}(\delta )=\hat{m}(\chi _{n})\ \ \ (\delta =1, \ldots ,12).
\label{med2}
\end{equation}

\subsubsection{NFDA applied to extreme value analysis}

\label{ext-val-app}Denote by $t_{\alpha }$ the $\alpha $-order quantile of
the distribution of $Y$ given a particular value of $\mathbf{\chi }$. From
the conditional distribution function, the $\alpha $-order quantile is
defined as:%
\begin{equation}
t_{\alpha }=F^{-1}(\alpha | \mathbf{\chi )},\ \ \ \ \forall \alpha \in (0,1).
\label{qant}
\end{equation}

Using the estimator given in (\ref{est-dist}), a nonparametric estimator of $%
t_{\alpha }$ in (\ref{qant}) is readily obtained by 
\begin{equation}
\hat{t}_{\alpha }=\hat{F}_{n}^{-1}(\alpha | \mathbf{\chi ).}  \label{qant-est}
\end{equation}

Several asymptotic properties of this estimator are shown in \cite%
{sankhya2005}. Expression (\ref{qant-est}) can be immediately used as an
estimator of the flow quantiles (\ref{rl}%
). Section~\ref{secondcase} presents an application of this approximation
using a time series of a river in the U.S.

The problem of extreme quantile estimation using functional data has also
been addressed in \cite{gardes}, where nonparametric estimators of quantiles
from heavy-tailed distributions when functional covariate information is
available are studied.

\section{Hydrological data}

\label{application}

In this Section, the functional nonparametric techniques are
applied to two time series of river flow (%
measured in cubic meters per
second, $m^3/s$), in the U.S., which were downloaded from 
the National Water Information System (NWIS) of USA,
http://waterdata.usgs.gov. 
The free statistical software R \citep{Rsoft} was employed to implement the different procedures.
Specific packages used in this process are cited below.

Firstly, flow data of Salt River
near Roosevelt, AZ, were selected. The annual peak flow data for this river were considered
by \cite{Kat02}, where they used a GEV distribution. A study
is also available in \cite{Anderson98}, who found that the monthly mean flow
is quite seasonal and possesses a heavy-tailed distribution. These data have
been also used in nonparametric studies \citep{Quintela2011}. In the present
paper, Salt River hydrological data are employed to examine the %
approaches on flow prediction and
extreme value analysis.
Additionally, a monthly mean flow time series of 
Christina River at Coochs Brigde, DE, was also considered \citep{usgs03, celebioglu}. These data
are only used in the time series prediction application, but not to perform extreme value analysis. Lower flow values,
compared with those of Salt River, are obtained here (see Figures \ref%
{fig1} and \ref{fig3}). 

\begin{figure}[h]
\centering
\includegraphics[width=10cm]{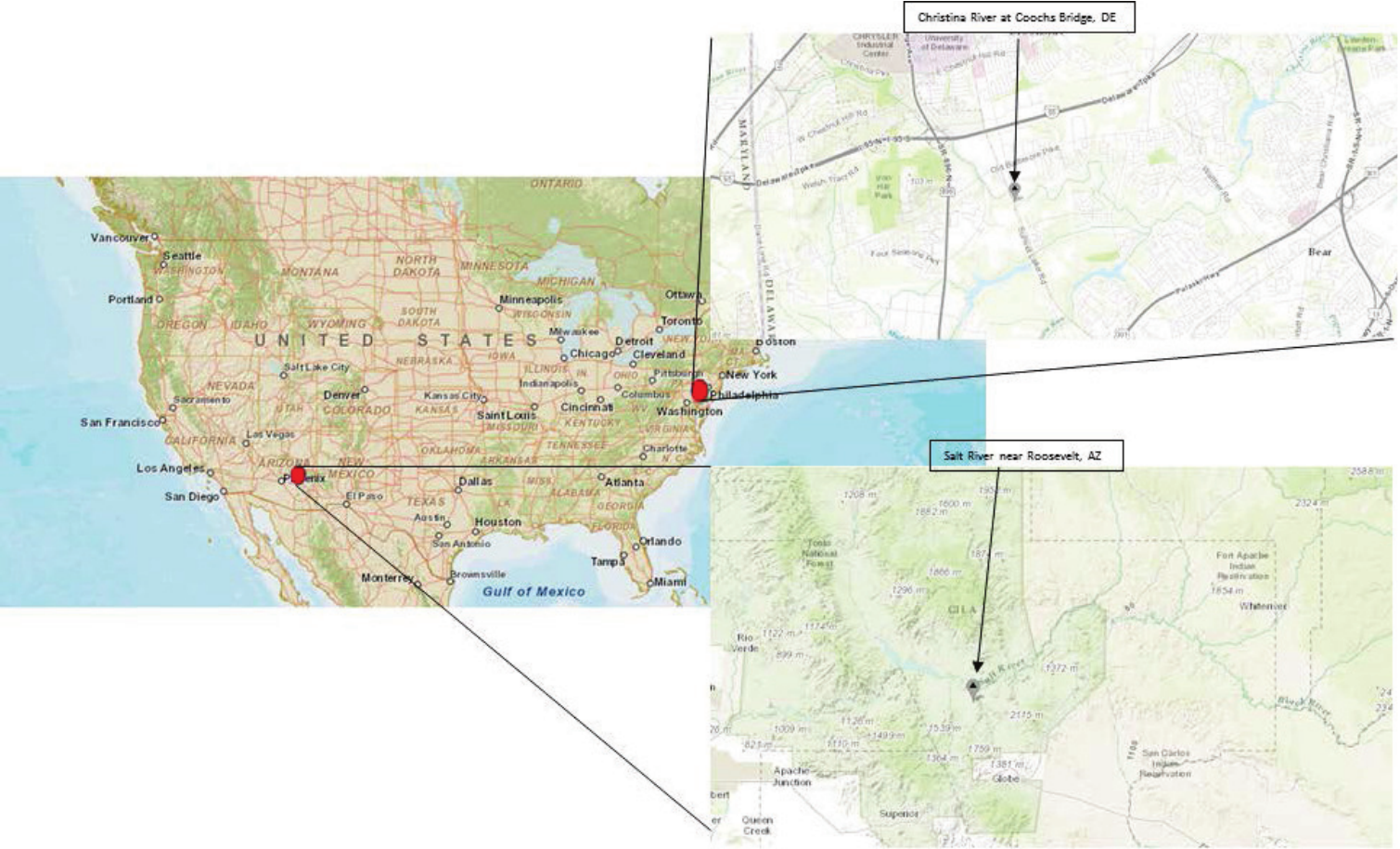}
\caption{Location map of Salt River near Roosevelt, AZ, and Christina River
at Coochs Brigde, DE.}
\label{map}
\end{figure}

These two rivers were selected because they belong to two different climate
areas with disparate temperatures and significant differences in rainfall
throughout the year (see Figure \ref{map} for a location map). Christina
River at Coochs Bridge at Delaware (US) is influenced by an Atlantic
climate, with high humidity and stable precipitations. The average annual
temperature in this location is about 13%
${{}^\circ}$%
C degrees, and the average annual precipitation is around 1168 mm. Salt
River near Roosevelt, Arizona, belongs to a Continental area, with high
average annual temperature (over 21%
${{}^\circ}$%
C degrees), and an average annual precipitation around 635 mm. Thus, the
performance of the NFDA techniques can be compared in different scenarios.

\subsection{Monthly mean flow prediction}

\label{firstcase}

Monthly mean flow data of both rivers, from January of 1944
to December of 2009, are considered.
The number of observations in this time interval is 792. In both cases, no missing values appear, and the quality
of the records is guaranteed by the information of the web page of the NWIS.

Firstly, a descriptive statistical analysis of both time
series is performed. Table \ref{dt1} presents the most usual descriptive
statistics for the data of the two rivers. In both cases, 
high values for the kurtosis and the skewness (to the right), and the
presence of maximum values far away from the rest of data, according to a
heavy-tailed distribution, can be observed.

\begin{table}[h]
\caption{Descriptive statistics for the monthly mean flow
variable of Salt River and Christina River.}
\label{dt1}
\begin{center}
\begin{tabular}{lcc}
\hline
Statistic & Salt River & Christina River \\
\hline
Minimum & 2.11 & 0.03 \\
1st Quartile & 5.88 & 0.32 \\
Median & 9.13 & 0.62 \\
Mean & 23.32 & 0.82 \\
3rd Quartile & 24.72 & 1.11 \\
Maximum & 381.47 & 4.68 \\
Standard Deviation & 35.93 & 0.67 \\
Skewness & 4.01 & 1.80 \\
Kurtosis & 23.13 & 7.85 \\
\hline
\end{tabular}%
\end{center}
\end{table}

\begin{figure}[h]
\centering
\includegraphics[width=10cm]{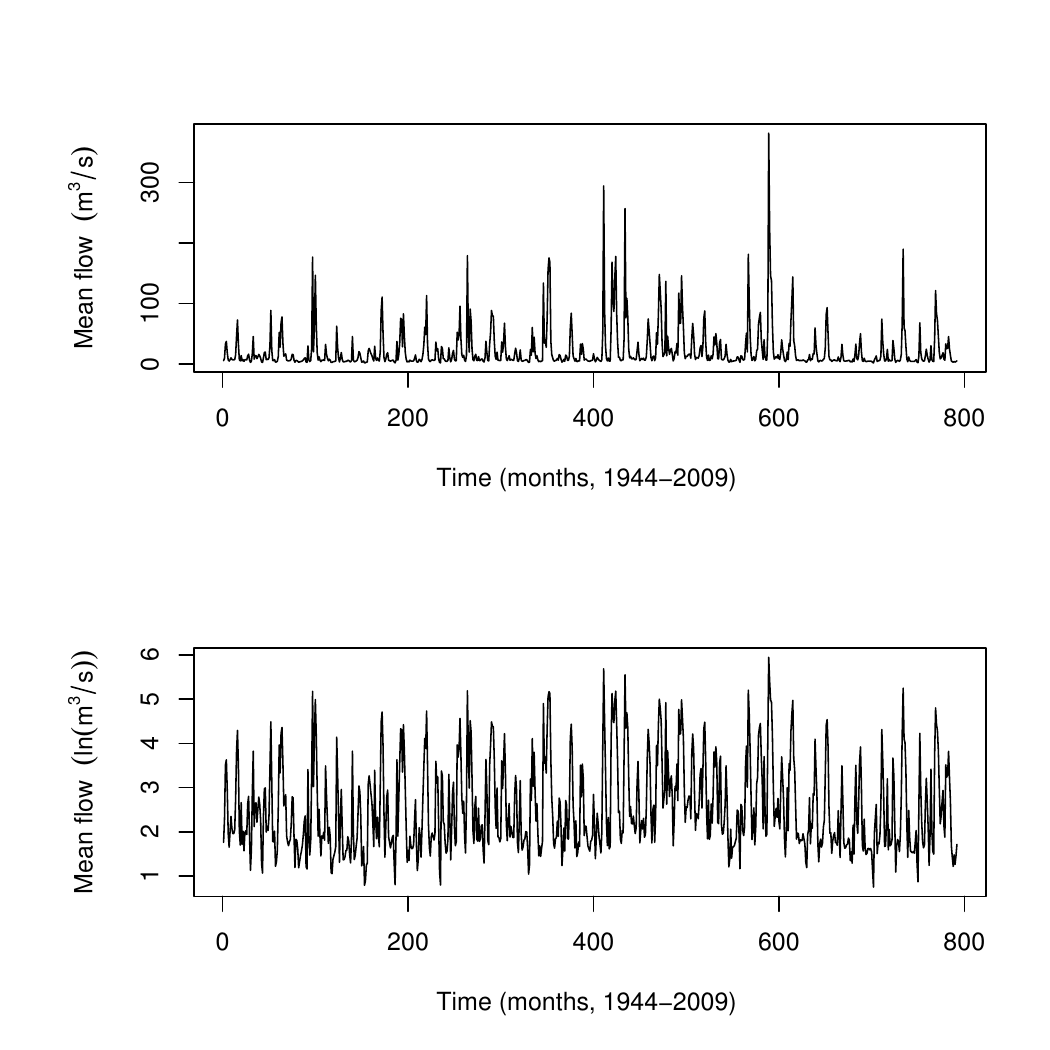}
\caption{Salt River monthly mean flow data. Top panel: original data (measured in $m^3/s$). Bottom panel: natural logarithm of original data.}
\label{fig1}
\end{figure}

\begin{figure}[h]
\centering
\includegraphics [width=10cm]{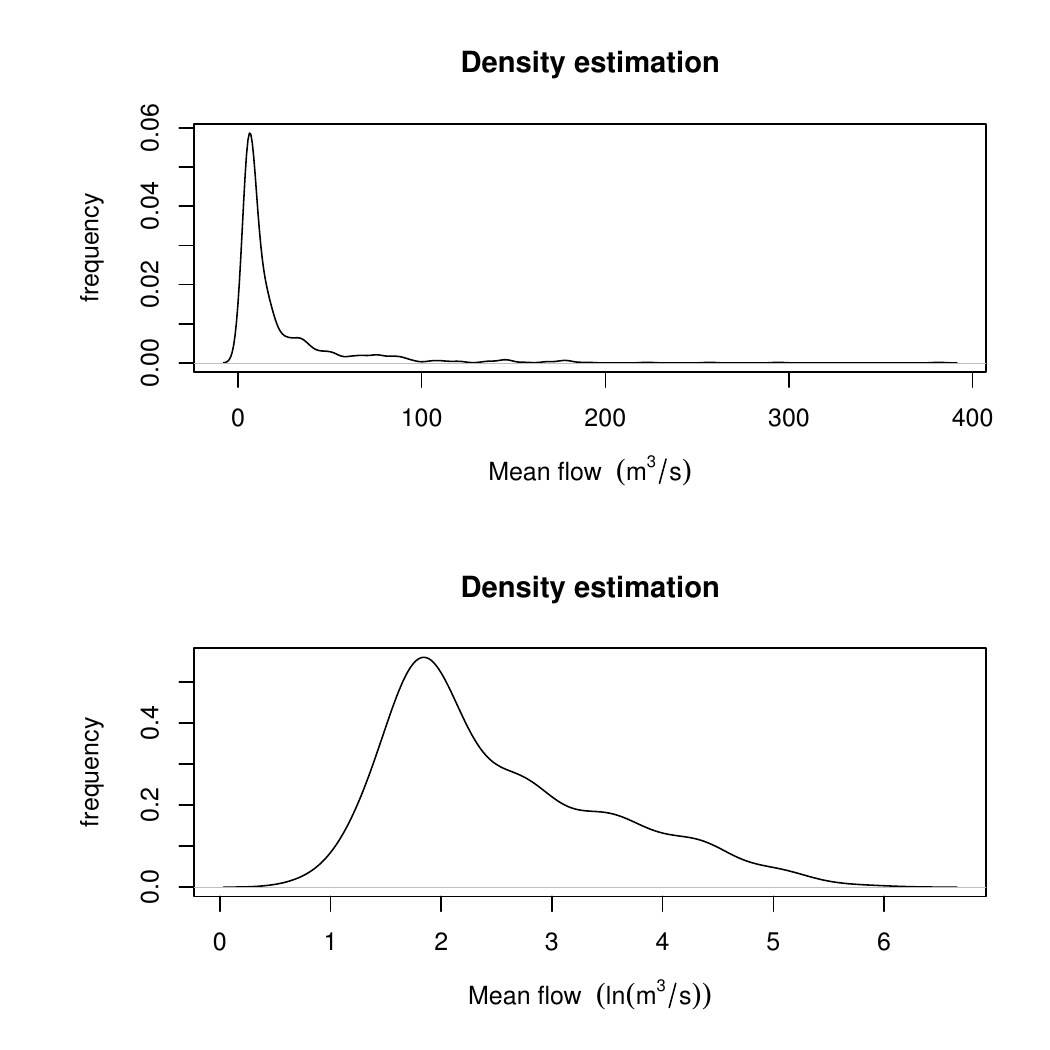}
\caption{Nonparametric density estimates of Salt River flow data. Top panel:
original data. Bottom panel: natural logarithm of original data.}
\label{fig2}
\end{figure}

\begin{figure}[h]
\centering
\includegraphics [width=10cm]{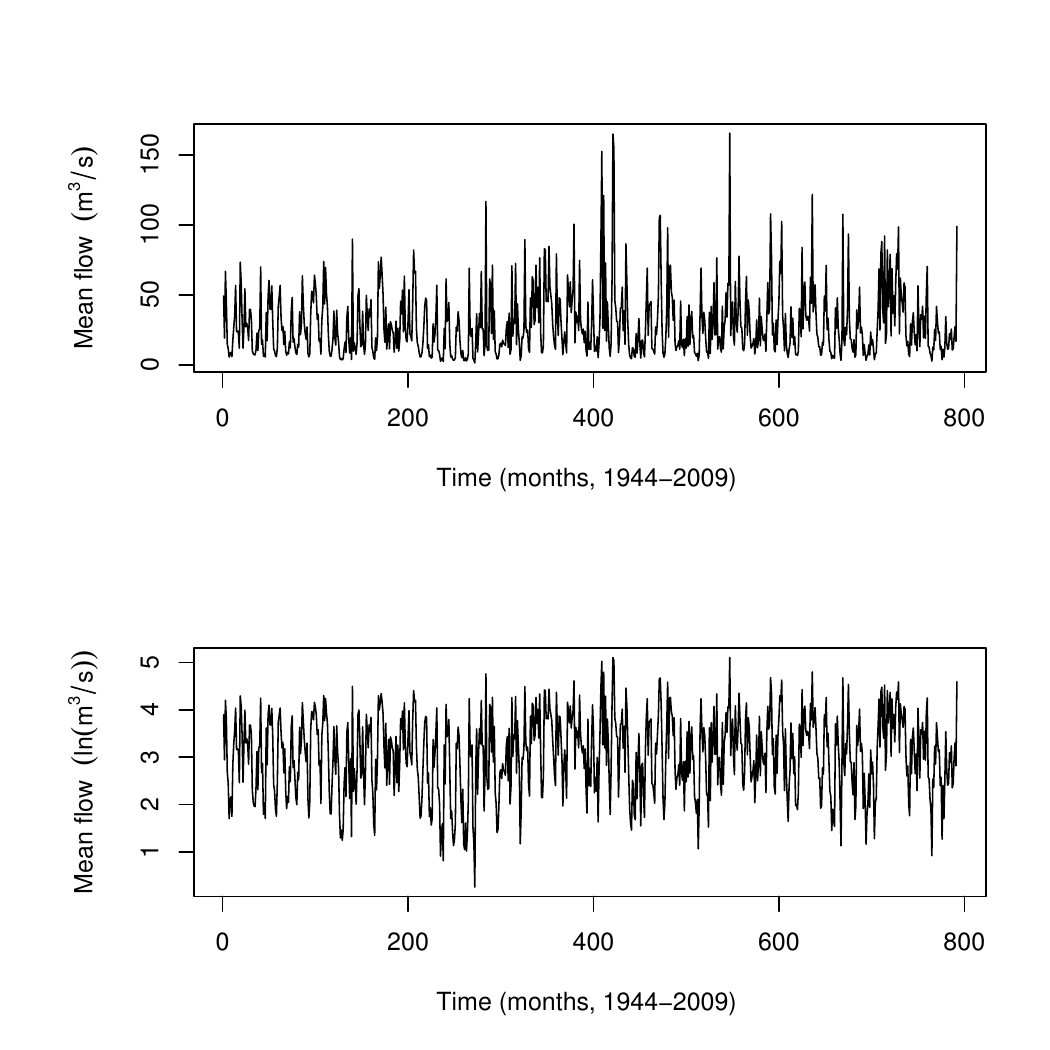}
\caption{Christina River monthly mean flow data. Top panel: original data (measured in $m^3/s$). Bottom panel: natural logarithm of original data.}
\label{fig3}
\end{figure}

The mean monthly time series, which does not fit a normal distribution, can
be normalised using a log-transformation function in order to remove the
periodicity of the original series \citep{Wang09,
Keskin06}. In Figure \ref{fig1}, Salt River data, before and after the
logarithmic transformation are shown. Figure \ref{fig2} presents the
estimated density functions computed with equation (\ref{kernel}) using
these data. In Figure \ref{fig3}, similar plots to those in Figure \ref{fig1}%
, but for Christina River, are displayed.

%
%

\begin{figure}[h]
\centering
\includegraphics[width=10cm]{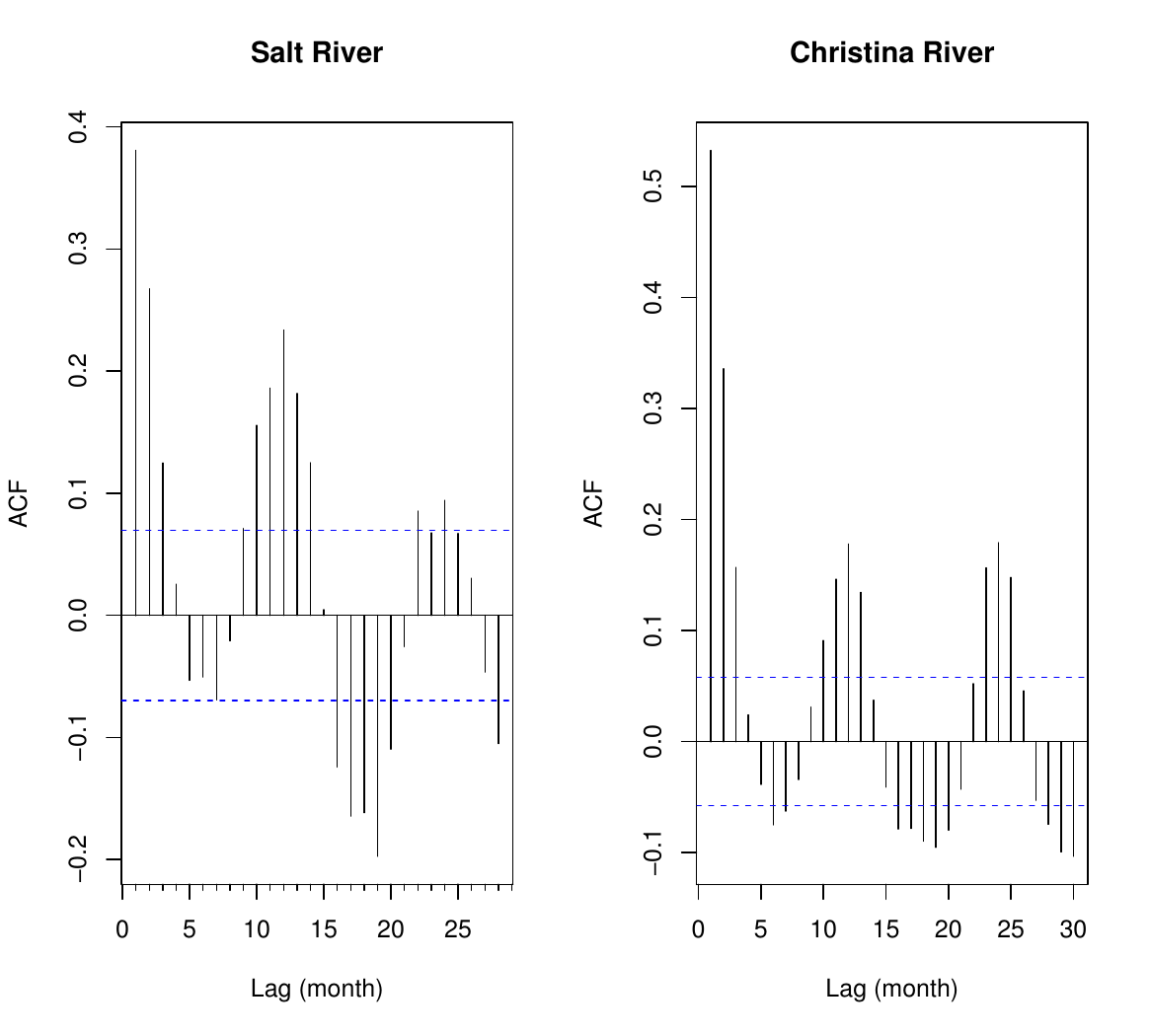}
\caption{Autocorrelation functions of Salt River and Christina River monthly mean flow data before the logarithmic transformation.}
\label{fig4-b-raw}
\end{figure}

\begin{figure}[h]
\centering
\includegraphics[width=10cm]{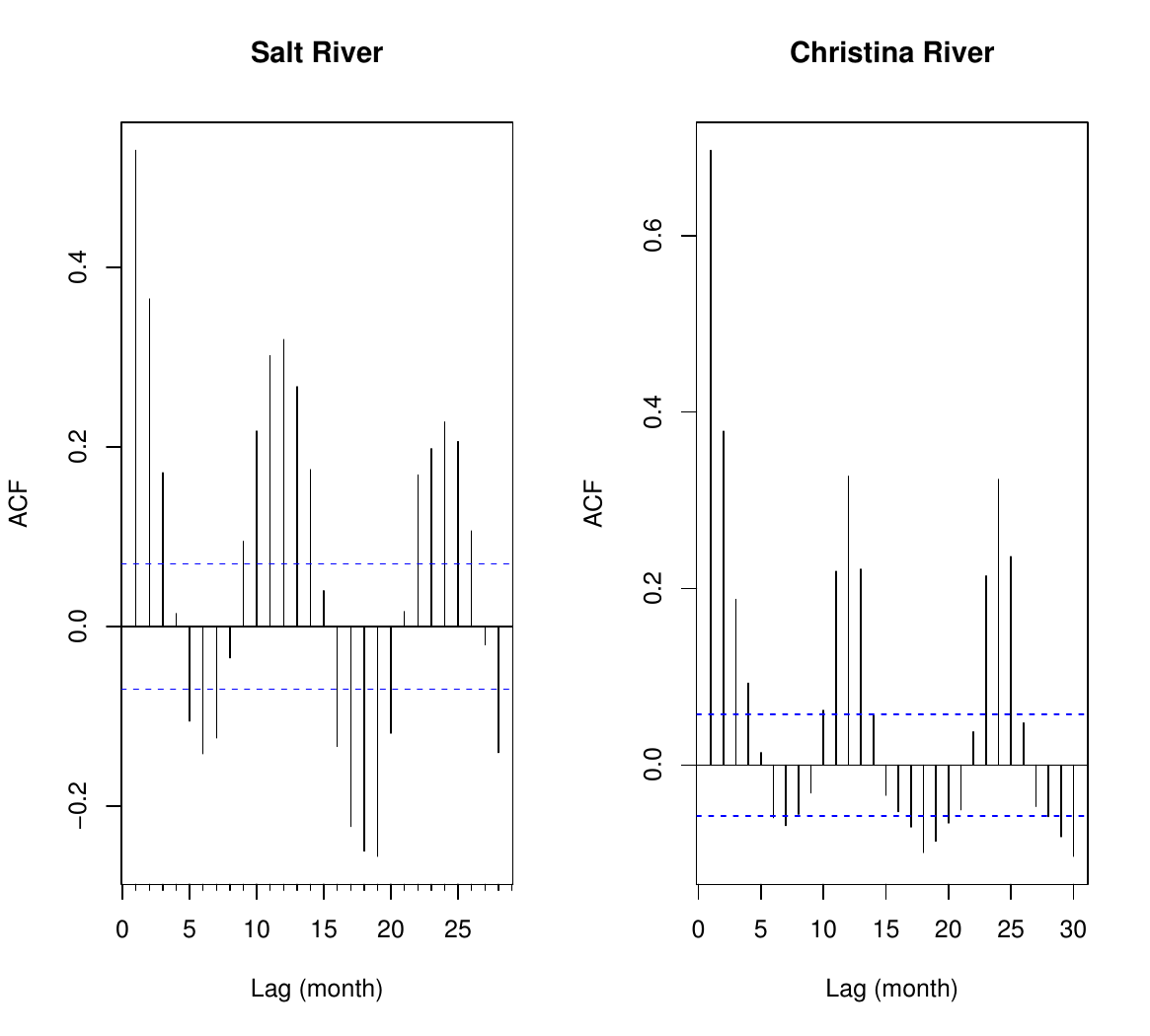}
\caption{Autocorrelation functions of Salt River and Christina River monthly mean flow data after the logarithmic transformation.}
\label{fig4-b}
\end{figure}

In Figures \ref{fig4-b-raw} and \ref{fig4-b}, the autocorrelation functions for the data of both rivers before and after the logarithmic transformation, respectively, are shown. The plots present different dependence structures, and suggest
that the $ARIMA$ modelling could be a possible approximation for prediction purposes.

To perform a functional analysis of the series and following Section \ref{pred-app}, the original
time series are converted into annual curves. In this case, there are no missing data and measures for all the months are available. Therefore, the number of annual curves is 66.
To validate the performance of the approaches, the values in the $66$th
year (2009) are predicted using the values from the $65$ previous years, and
these predictions are compared with the real values in that year. To apply the
nonparametric functional methods, two bandwidths have to be selected. To do
this, in a first step, considering the first $64$ years, the $65$th is used
as a validation step. 
As explained in Section \ref{pred-app}, given the sample $\{\chi _{i},Y_{i}(\delta
)\}_{i=1}^{64}$, a predictor of $%
Y_{64}(\delta )$, knowing $\chi _{64}$, can be achieved using equation (\ref{med1}).
Repeating this step for $\delta =1,\ldots
,12,$ the mean values of the flow for the $65$th year can be predicted. 
The NFDA estimators are applied using kernels based on the Epanechnikov
density, $G(u)=0.75(1-u^{2})1_{[-1,1]}(u)$, taking $K(u)=2G(u)1_{[0,1]}(u)$\
and $H(u)=\int_{-\infty }^{u}G(t)dt$. On the other hand, the bandwidths$\ h$
and $g$ are selected by minimizing the prediction error over the $65$th
year, that is $\sum_{\delta =1}^{12}(\hat{Y}_{64}(\delta )-Y_{64}(\delta
))^{2},$ and the FPCA (Functional Principal Components Analysis) semi-metric
is used \citep[for more details,
see][]{ramsaysilvermanlibro}.

Next, in a second step, given $\{\chi _{i},Y_{i}(\delta )\}_{i=1}^{64}$ and
the previous selected parameters $h$ and $g $, $\hat{F}_{n}(\cdot | \chi
_{65})$ is estimated and a predictor of $Y_{65}(\delta )$ for $\delta
=1,\ldots ,12$, using the corresponding estimator of the median of the
conditional distribution $F(\cdot | \chi _{65})$ given in (\ref{med1}), is obtained. Additionally, the nonparametric functional estimator of the mean function (\ref{noparareg}), based on
regression techniques, was also applied. In this case, the monthly mean flows of the $66$th year were predicted using equation (\ref{med2}) for $n=65$.
The software
for computing the NFDA, programmed in R, can be freely obtained at the web
http://www.math.univ-toulouse.fr/staph/npfda/.

A parametric $ARIMA$ model is also fitted to the time series, by means of
the package \texttt{forecast} of the software R. In this package, automatic
methods to select the order of the model and also to estimate the
corresponding parameters are implemented. In this case, an $ARIMA(1,0,2)$ is
fitted for Salt River data and an $ARIMA(4,0,4)$ for
Christina River data.

\begin{figure}[h]
\centering
\includegraphics [width=10cm]{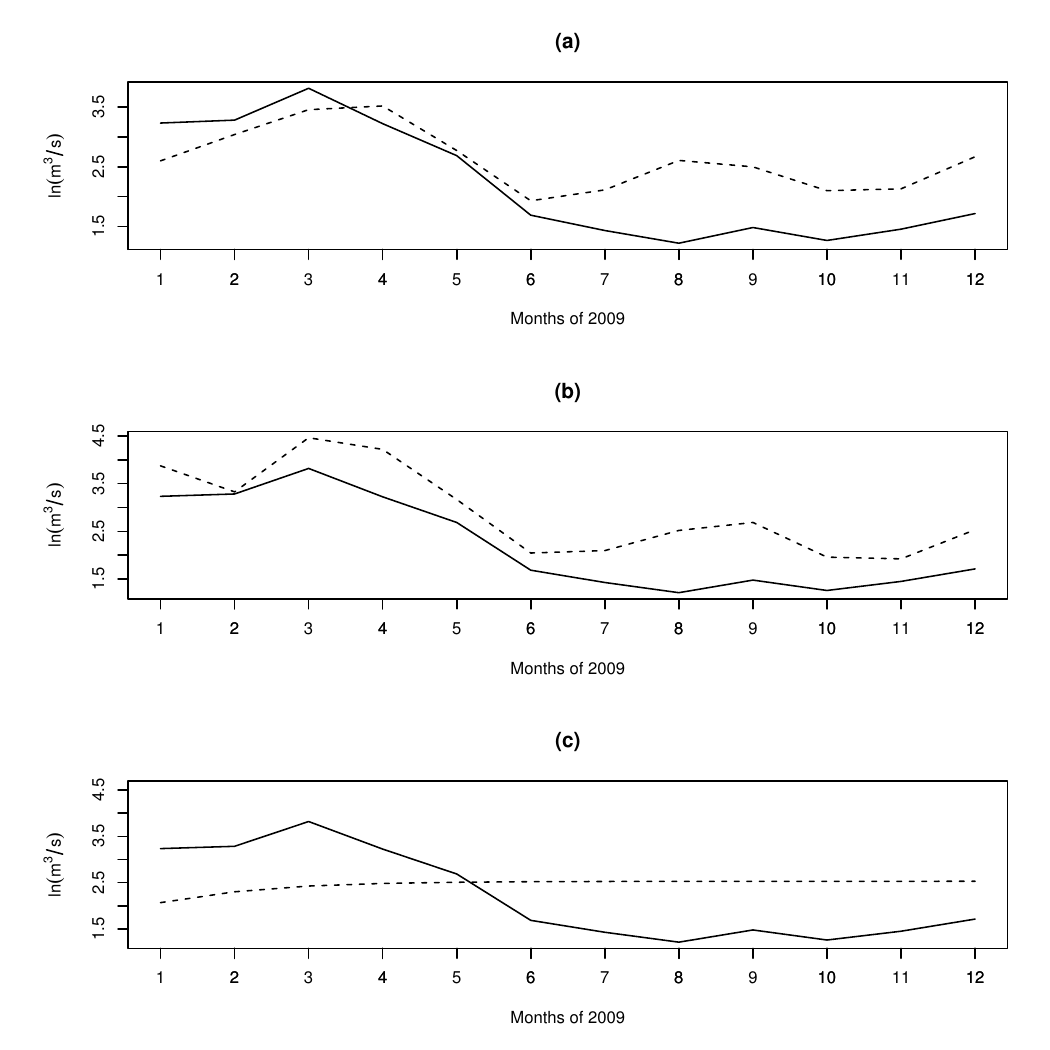}
\caption{Predicted values for Salt River monthly mean flows in
2009 (dashed line) and real values in that year (solid line). (a)
nonparametric functional data (NFDA) modelling based on regression. (b)
nonparametric functional data (NFDA) modelling based on the median. (c) $%
ARIMA $ approach.}
\label{fig4}
\end{figure}

\begin{figure}[h]
\centering
\includegraphics [width=10cm]{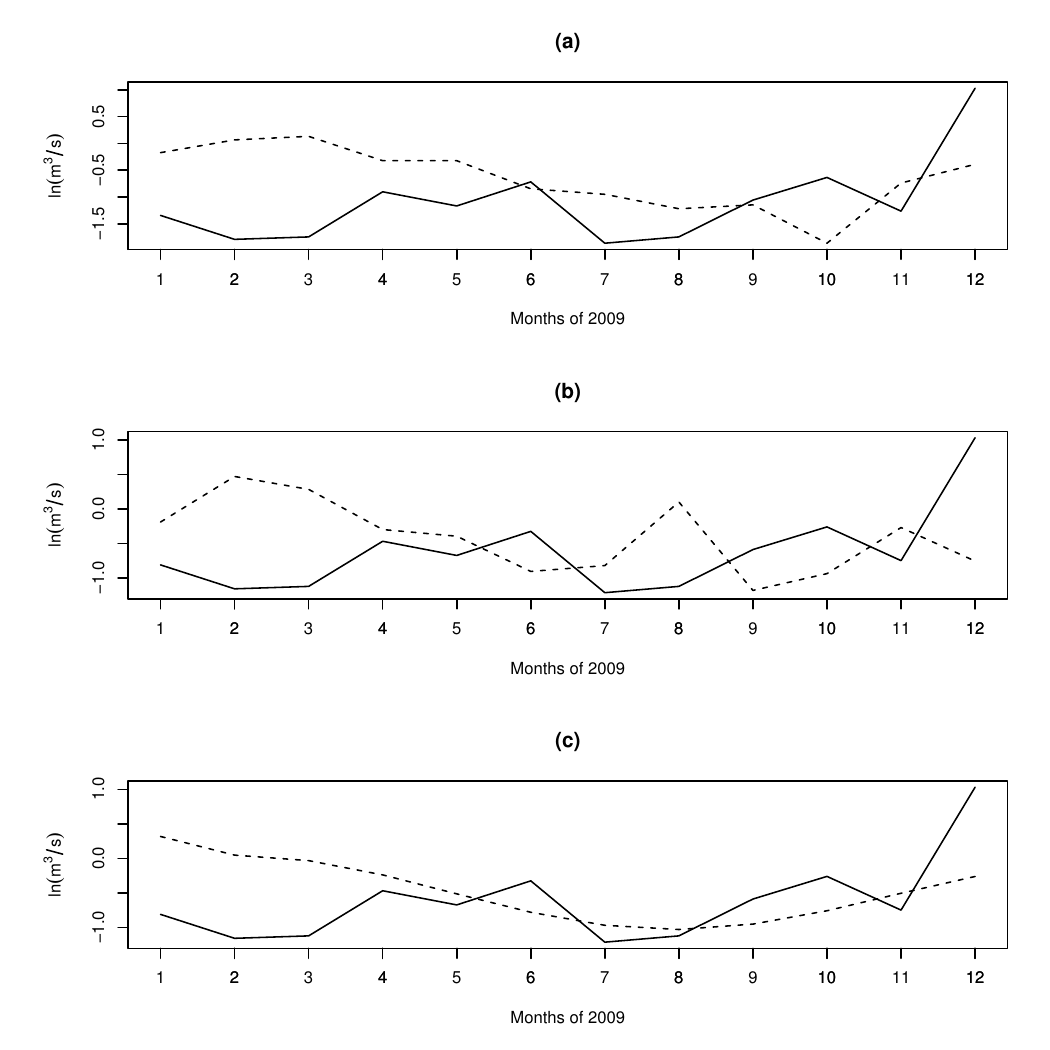}
\caption{Predicted values for Christina River monthly mean flows
in 2009 (dashed line) and real values in that year (solid line). (a)
nonparametric functional data (NFDA) modelling based on regression. (b)
nonparametric functional data (NFDA) modelling based on the median. (c) $%
ARIMA $ approach.}
\label{fig5}
\end{figure}

Figure \ref{fig4} shows, for Salt River data, the predicted values for
the $66$th year (dashed line) together with the real values in the $66$th
year (solid line). All the data considered are the natural logarithm of the real
values. Figure \ref{fig5} is the equivalent plot for the Christina River
data set. In each case, the top panel corresponds to the functional modelling
using the predictions based on regression (equation (\ref{med2})), the
middle panel shows the functional modelling using the predictions based on the
median (equation (\ref{med1})), and the bottom panel presents the $ARIMA$
approach.

%

A numerical comparison for obtaining the best predictor is made using the
MSE criterion, that is, 
\begin{equation}
\mathrm{MSE}=\frac{1}{12}\sum_{\delta =1}^{12}(\hat{Y}_{65}(\delta
)-Y_{65}(\delta ))^{2}.  \label{mse}
\end{equation}

The MSE values in both rivers are given in Table \ref{t1}. In the first
row, the results using NFDA based on regression (equation (\ref{med2})) are presented. The results obtained applying NFDA methods based on the median (equation (\ref{med1})) are shown in the second row. Finally, in the third row, MSE values using $ARIMA$ models are given.


\begin{table}[h]
\caption{MSEs of the monthly mean flow predictors in the $66$th year
(2009) using different methods (NFDA based on regression, in the first
row; NFDA based on the median, in the second row; and $ARIMA$ models in the
third row), for Salt River and Christina River.}
\label{t1}
\begin{center}
\begin{tabular}{l|cc}
\hline
& \multicolumn{2}{|c}{River} \\ 
Method & Salt River & Christina River \\ \hline
NFDA based on regression & $0.5965$ & $0.7388$ \\ 
NFDA based on the median & $0.5208$ & $0.4969$ \\ 
$ARIMA$ models & $1.0818$ & $0.9430$ \\ \hline
\end{tabular}%
\end{center}
\end{table}

As observed in Figures \ref{fig4} and \ref{fig5}, NFDA predictions methods
provide better fits to the real series. The $ARIMA$ predictions are,
basically, the mean values. Moreover, it can be observed in Table \ref{t1}
that the MSE errors are lower using the NFDA techniques, and the best
criterion is that using the median as the predicted value in the two time
series.

\subsection{Extreme value analysis}

\label{secondcase}

\begin{figure}[h]
\centering
\includegraphics [width=10cm]{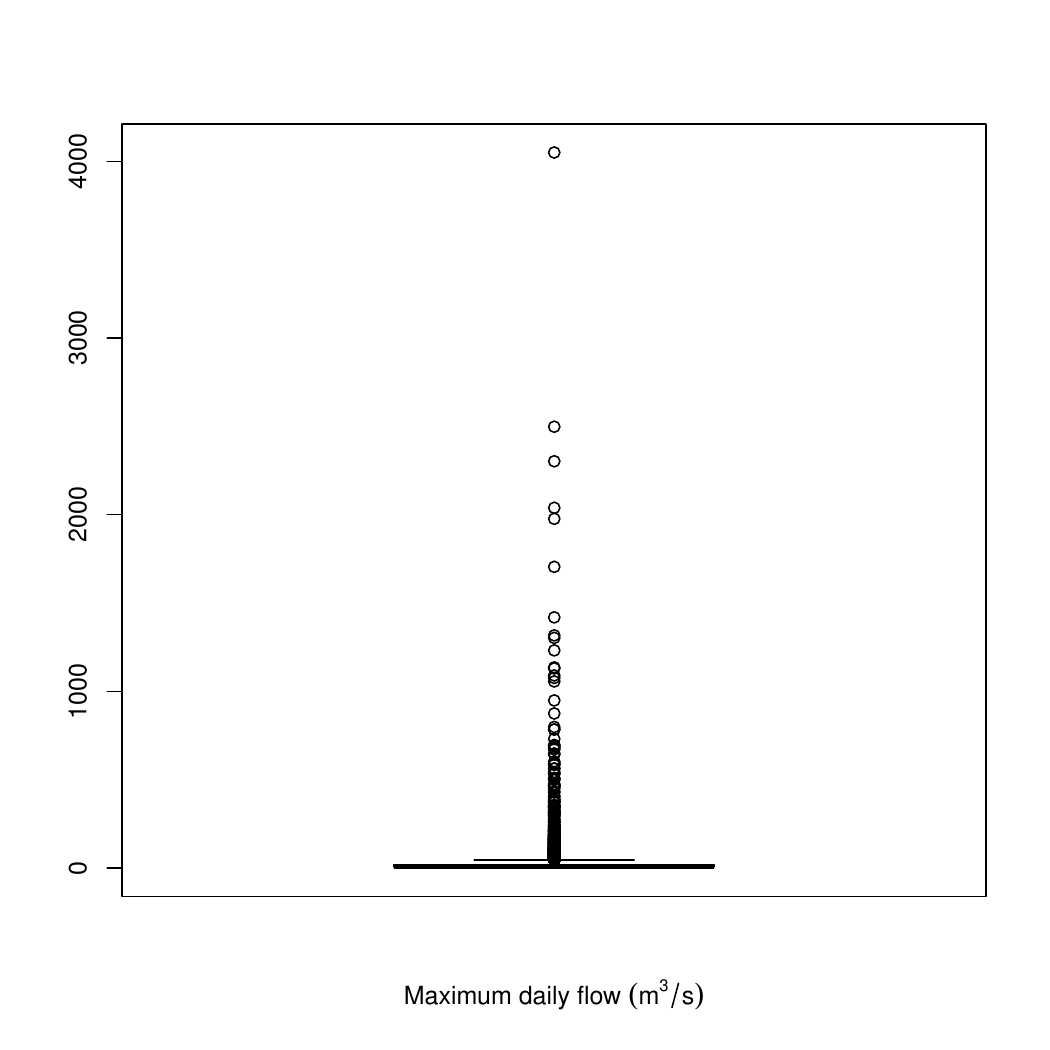}
\caption{Boxplot of Salt River maximum daily flow data, measured in
$m^3/s$.}
\label{fig6}
\end{figure}

In this section, NFDA techniques are applied for extreme value analysis.
Equations described in Section \ref{ext-val-app} are used, and the results
obtained are compared with those using the parametric  GEV and
nonparametric estimators presented in Sections \ref{ext-val-an} and \ref%
{nopara}, respectively. In this case, only Salt River data are available.
The maximum daily flow data of this river, from 01/01/1987 to 31/12/2009,
are used to calculate flow quantile estimates as indicated in (\ref{rl}). In
Figure \ref{fig6}, 
a boxplot computed with these data is presented.
It can be observed the very asymmetric and
heavy-tailed data distribution, with a lot of extreme values corresponding to high quantiles of the variable. Similar information can be deduced from
Table \ref{t2}, where the most usual descriptive statistics for the maximum
daily flow variable are shown.

%

\begin{table}[h]
\caption{Descriptive statistics for Salt River maximum daily flow
variable.}
\label{t2}
\begin{center}
\begin{tabular}{lr}
\hline
Minimum & 1.95 \\
1st Quartile & 5.55 \\
Median & 8.26 \\
Mean & 27.25 \\
3rd Quartile & 21.42 \\
Maximum & 4050.00 \\
Standard Deviation & 94.51 \\
Skewness & 19.63 \\
Kurtosis & 592.33 \\
\hline
\end{tabular}%
\end{center}
\end{table}

The considered values from years 1987-2008 (inclusive) are used in the
estimation process, and the corresponding estimates are checked with the
real values in the year 2009.

In the classical parametric GEV estimation (Section \ref%
{ext-val-an}), the data need to be independent, or, at least, the
dependence has to decrease suitably fast with increasing time separation %
\citep{Smith89}. However, nonparametric estimators (both of functional and
non-functional type) can be correctly applied in this field and have good
theoretical properties, although the assumption of independence is not
strictly fulfilled \citep{Youndj06,quintela2008}.

The first step to apply NFDA (now, to calculate flow quantiles) is to
construct functional data from a sample of daily maxima, in the same way as
in Section \ref{firstcase}. Because daily data are available,
functional data composed of the corresponding values of each month are constructed. Unlike
the situation in Section \ref{firstcase}, now the number of components
changes from one functional variable $\chi $ to another (unbalanced data
setting). This happens because the months do not have the same number of
days. All months are considered to have 31 measures, interpolating
linearly the two closest values for each value that originally does not
exist. Therefore, each functional observation consists of 31 data.

Here, the construction of the functional data is analogous to (\ref{f1}):
\begin{equation}
\forall t\in \{1,2,\ldots,31\},\ \ \ \ \ \chi_{i}(t)=Z_{31\cdot (i-1)+t},
\label{f2}
\end{equation}
where $\{Z_{k}\}_{k=1}^{n}$ denotes the complete time series of daily
maxima, and $\chi _{i}$ $=(\chi _{i}(1),\ldots ,\chi _{i}(31))$ the daily
data of the $i$th month. Now, our focus is on the estimation of the
conditional distribution function of the variable of each daily maximum,
conditioned on the values in the previous month.

The comparison between the classical parametric GEV methods, the nonparametric
techniques and the NFDA approaches is carried out in the following steps: a
set of values for levels $c_{i}$ from$\ i=1,\ldots ,20$ is selected.
Specifically, $c_{1}$ is chosen as the median of the data (up to the year
2008), and $c_{20}$ as the quantile of order 0.95. The sequence $c_{i}$
consists of 20 equally spaced points. Using the true measures of the last
year 2009, the number of days in which the values $c_{i}$ were exceeded can
be computed. Thus, the recurrence intervals, using the corresponding
empirical distribution function, 
\begin{equation*}
F_{n}(c_i)=\frac{\mathrm{Number \ of \ measures \ } \leq c_{i}}{365},
\end{equation*}
in expression (\ref{mrp}), can be approximated. These estimators are: 
\begin{equation}
\hat{RT}(c_{i})=\frac{1}{1-F_{n}(c_{i})},\ \ i=1,\ldots,20.  \label{tevel}
\end{equation}

Now, any estimation method of the flow quantiles (\ref{rl}), using the
values in (\ref{tevel}), should provide an approximated value of the true
values $c_{i}$. The flow quantiles are estimated using the classical
parametric methods, the nonparametric approaches and also by means of our
approximation based on NFDA methods, described below.

\begin{description}
\item[Parametric GEV approach.] For each $i=1,\ldots ,20$, the flow
quantiles are estimated by 
\begin{equation}
\hat{c_{i}}_{\theta }=F_{\hat{\theta}}^{-1}\left( 1-\frac{1}{\hat{RT}(c_{i})}%
\right) .  \label{rlpara}
\end{equation}%
For this, the package \texttt{evir} of the software R, that estimates
the GEV parameters by maximum likelihood, is used.

\item[Nonparametric approach.] For each $i=1,\ldots ,20$, nonparametric
estimators of the flow quantiles are calculated, as indicated in (\ref%
{rlnopara}):%
\begin{equation*}
\hat{c}_{ih}=F_{h}^{-1}\left( 1-\frac{1}{\hat{RT}(c_{i})}\right) ,
\end{equation*}%
where the bandwidth, obtained by cross-validation, is $h=13.66$

\item[NFDA approach.] Expression (\ref{med1}) can be adapted to estimate any
quantile. Then, for each $i=1,\ldots ,20$, estimators of the flow quantiles
are obtained by estimating the conditional quantile of order $1/\hat{RT}%
(c_{i})$ by the expression: 
\begin{equation}
\hat{Y}_{(n/31)-1}(\delta )=\hat{F}_{n}^{-1}\left( \left. 1-\frac{1}{\hat{RT}%
(c_{i})}\right| \chi _{(n/31)-1}\right) ,\ \ \delta =1,2,\ldots ,31,
\label{1.2}
\end{equation}%
where, for each $\delta $, $Y_{j}(\delta )=\chi _{j+1}(\delta )$ and $\chi
_{(n/31)-1}$ denotes the functional data composed of the 31 measures of the
penultimate month. Then, for each day $\delta $ an estimated value is available,
and the functional nonparametric estimate of the flow quantile, denoted by $%
\hat{c}_{iF}$, will be the sample mean of these daily values $\hat{Y}%
_{(n/31)-1}(\delta )$. The same kernels, bandwidths and semi-metric as in
the example in Section \ref{firstcase} are used.
\end{description}

\begin{figure}[h]
\centering
\includegraphics [width=10cm]{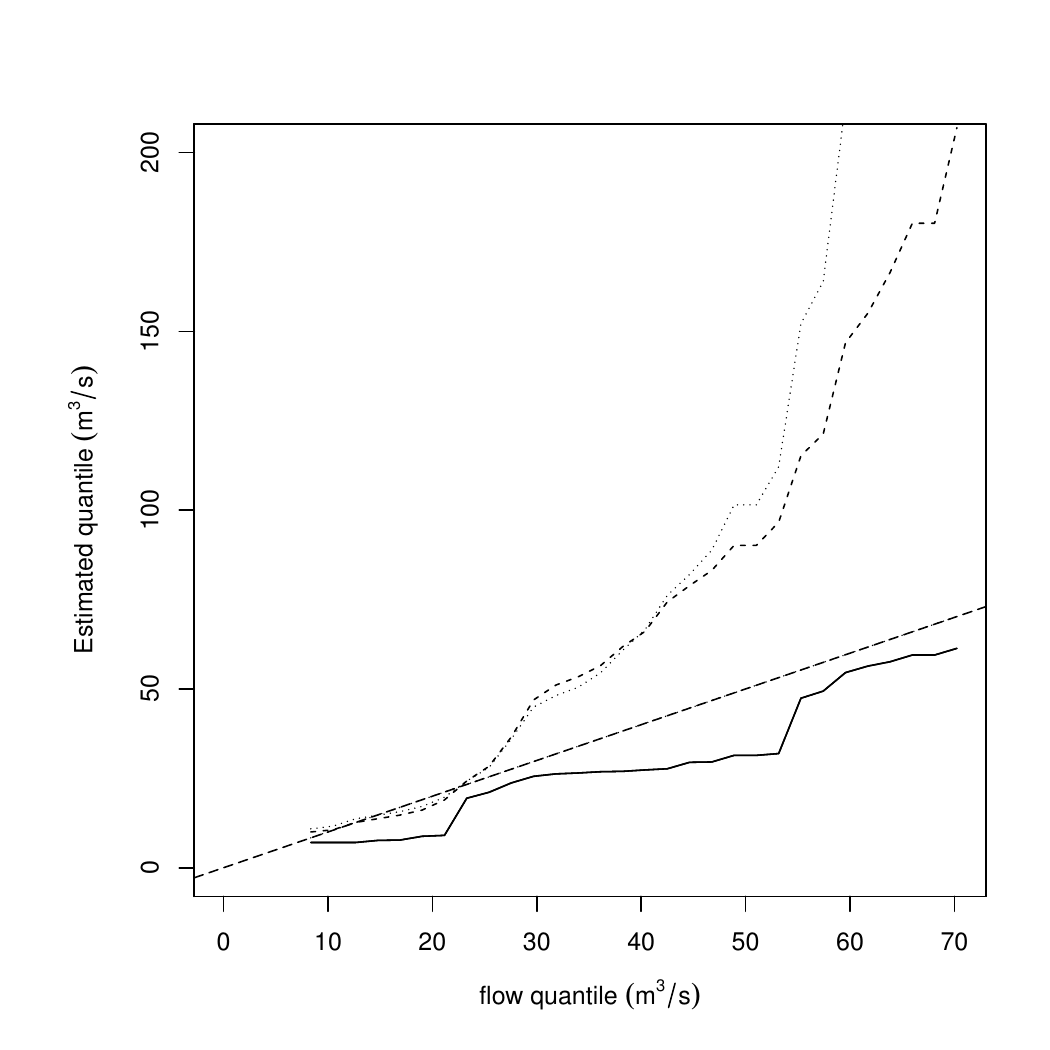}
\caption{Estimations of the quantiles using the parametric GEV estimator,
nonparametric kernel method and the nonparametric functional (NFDA) approach
for Salt River data (parametric GEV estimations with a dotted line,
nonparametric with a dashed line and nonparametric functional estimations
with a solid line. The dashed diagonal line represents the true quantiles to
be estimated.}
\label{fig7}
\end{figure}

To compare mathematically the three approaches, the relative mean absolute
error (RMAE) of $\hat{c_{i}}_{\theta }$, $\hat{c}_{ih}$ and $\hat{c}_{iF}$
is computed, given by: 
\begin{equation}
\mathrm{RMAE}=\frac{1}{20}\sum_{i=1}^{20}\frac{\left\vert c_{i}-\hat{c}%
_{i\ast }\right\vert }{c_{i}},  \label{rmae}
\end{equation}%
where $\hat{c}_{i\ast }$ can be $\hat{c_{i}}_{\theta }$, $\hat{c}_{ih}$ or $%
\hat{c}_{iF}$. The results obtained are%
\begin{equation}
\mathrm{RMAE}=1.13,0.71\ \text{and }0.26,  \label{rmae2}
\end{equation}%
for the parametric GEV, nonparametric and NFDA estimates, respectively.
Therefore, the minimum error is obtained with the NFDA techniques. On
the other hand, Figure \ref{fig7} shows the quantile estimations with the
previous proposals for Salt River data (parametric GEV estimations with a
dotted line, nonparametric with a dashed line, and NFDA
estimations with a solid line. The dashed diagonal line represents the true
values to be estimated). 
The long-tailed distribution observed in Figure \ref{fig6} clearly reveals
the difficulty in the extreme value estimation process \citep{Serinaldi09}.
However, the NFDA approach, considering each functional datum as the
complete set of values for each month, gives more precise estimations than
those obtained with the parametric GEV or the simple nonparametric methods. The
largest differences between the estimates occur at the highest levels, where
the good approximations of the NFDA estimates are observed and it is more
important to have reliable prediction techniques. Note that a multivariate
approach would be possible in the parametric GEV and the nonparametric settings,
but, in this case, a vector composed of 30 predictor variables would be
necessary. This high value makes very difficult (if not impossible) this
kind of approximation in practice.

\section{Discussion}

\label{conclussions}

Statistical techniques are usually applied to address practical problems in
hydrology. In the present paper, two of them, monthly mean river
flow prediction and extreme value analysis, are the focus of the research. NFDA approaches, combining
nonparametric methods with functional data, are used in this setting.

The main objective of this paper was to apply different NFDA
techniques to two particular hydrological problems, and to test their
behaviour in comparison to more classical approaches. The nonparametric
functional methods were applied to real data of two rivers in the U.S. The
different alternatives were validated using the final year in the database
as a testing sample, and the rest of the years as the training sample.

In the prediction setting, two nonparametric functional proposals, based on
the median and the mean, respectively, were applied and compared with
classical $ARIMA$ models. The results showed that NFDA approaches, especially
those based on the median, had a better performance than the classical $ARIMA$
models.

The previous approaches could be extended including available information
like daily precipitation, daily temperature or any other climatic covariate.
Several models similar to those presented incorporating covariates have been
proposed and studied previously. For example, the dynamic regression
($ARIMAX$) combines the Box-Jenkins models with the linear regression,
obtaining a more general model for the study of the time series %
\citep{Shumway}. This kind of models simply adds covariates to the general
expression of an $ARIMA$ model, but the covariate coefficients are hard to
interpret. An alternative approach could be the application of regression
models with $ARMA$ (or $ARIMA$) errors. This includes the use of parametric,
nonparametric or semiparametric approaches. In a hydrological context, \cite%
{castellano04} presents a study 
of the Xallas
river (northwest of Spain), using Box-Jenkins and neural networks methods,
incorporating exogenous variables such as rainfall information. Regarding
the case of functional methods, covariates could be included in the problem
through the use of semi-functional partial linear models \citep{aneiros06}.
This approach uses a nonparametric kernel procedure; the output is scalar,
and a functional covariate and multivariate non functional covariate are
considered. Functional regression between functional explanatory variables
and a scalar response is also possible using the backfitting algorithm %
\citep{febrero11}. This would allow including functional covariates in the
model. The application of these techniques to our data would require the
availability of some relevant climatic variables. Unfortunately, these
variables are not available in the managed databases. However, a
more deep study of this issue could be carried out in a future research.

Regarding the extreme value analysis,  the estimation of the 
flow quantiles has been the focus of the study. These values play an important role in hydrological
problems, because they are directly linked with flood analysis. 
The new NFDA approach performed better than the parametric
GEV estimators, 
producing more close estimations to the
true values. On the other hand, it is well-known that due to the small
number of extreme values in a sample, it is usually difficult to obtain
reliable estimations. These estimations could be
improved by using more precise bandwidth parameters. The bandwidth parameter
selection in NFDA remains, nowadays, as an open problem. The development of
data-driven techniques for computing optimal bandwidths will produce
directly the improvement of the promising results obtained in the
quantile estimation problem.

In general, the approaches proposed in this paper yielded accurate
estimates of both the functions of interest, such as the cumulative distribution function or the function providing the probabilities of exceedance,  and derived parameters, as, for example, the flow quantiles. They also captured
more complex patterns in the data providing better future estimations.
Therefore, they represent a better alternative to the classical methods
regularly used in this framework, being useful tools for environmental
agencies 
to manage hydrological risks including those of floods.

\section{Acknowledgements}

The authors have no conflict of interest to declare. This research has been
partially supported by the Spanish Ministry of Science and Innovation Grants
MTM2011-22392 and MTM2014-52876-R for the second author.
We are grateful to three referees and to the editors of the Journal of Flood Risk Management, for constructive and helpful suggestions.


\begin{thebibliography}{55}
\expandafter\ifx\csname natexlab\endcsname\relax\def\natexlab#1{#1}\fi
\providecommand{\bibinfo}[2]{#2}
\ifx\xfnm\undefined \def\xfnm[#1]{\unskip,\space#1}\fi
\bibitem[{Adamowski(1989)}]{Adamowski89}
\bibinfo{author}{Adamowski\xfnm[ K.]}.
 \bibinfo{year}{1989.} \newblock \bibinfo{title}{A {M}onte {C}arlo comparison of parametric and
  nonparametric estimation of flood frequencies}.
\newblock \bibinfo{journal}{{\it Journal of Hydrology}}
\bibinfo{volume}{{\bf 108}} : \bibinfo{pages}{295--308}.
\bibitem[{Adamowski and Feluch(1990)}]{AdamowskiFeluch90}
\bibinfo{author}{Adamowski\xfnm[ K.]}, \bibinfo{author}{Feluch\xfnm[ W.]}.
\bibinfo{year}{1990.} \newblock \bibinfo{title}{Nonparametric flood frequency analysis with
  historical information}.
\newblock \bibinfo{journal}{{\it Journal of Hydraulic Engineering}}
 \bibinfo{volume}{{\bf 116}} : 1035--1047.
\bibitem[{Anderson and Meerschaert(1998)}]{Anderson98}
\bibinfo{author}{Anderson\xfnm[ P.L.]}, \bibinfo{author}{Meerschaert\xfnm[
  M.M.]}.
\bibinfo{year}{1998.} \newblock \bibinfo{title}{Modeling river flows with heavy tails}.
\newblock \bibinfo{journal}{{\it Water Resources Research}}
 \bibinfo{volume}{{\bf 34}} : 2271--2280.
\bibitem[{Aneiros-Perez and Vieu(2006)}]{aneiros06}
\bibinfo{author}{Aneiros-Perez\xfnm[ G.]}, \bibinfo{author}{Vieu\xfnm[ P.]}.
  \bibinfo{year}{2006.} \newblock \bibinfo{title}{Semi-functional partial linear regression}.
\newblock \bibinfo{journal}{{\it Statistics \& Probability Letters}}
\bibinfo{volume}{{\bf 76}} : 1102--1110.
\bibitem[{Apipattanavis et~al.(2010)Apipattanavis, Rajagopalan and
  Lall}]{apipa}
\bibinfo{author}{Apipattanavis\xfnm[ S.]}, \bibinfo{author}{Rajagopalan\xfnm[
  B.]}, \bibinfo{author}{Lall\xfnm[ U.]}.
\bibinfo{year}{2010.} \newblock \bibinfo{title}{Local polynomial--based flood frequency estimator for
  mixed population}.
\newblock \bibinfo{journal}{{\it Journal of Hydraulic Engineering}}
 \bibinfo{volume}{{\bf 15}} : 680--691.
\bibitem[{Besse et~al.(2000)Besse, Cardot and Stephenson}]{besse2000}
\bibinfo{author}{Besse\xfnm[ P.]}, \bibinfo{author}{Cardot\xfnm[ H.]},
  \bibinfo{author}{Stephenson\xfnm[ D.]}.
\bibinfo{year}{2000.} \newblock \bibinfo{title}{Autoregressive forecasting of some functional
  climatic variations}.
\newblock \bibinfo{journal}{{\it Scandinavian Journal of Statistics}}
\bibinfo{volume}{{\bf 27}} : 673--687.
\bibitem[{Bowman et~al.(1998)Bowman, Hall and Prvan}]{Bowman98}
\bibinfo{author}{Bowman\xfnm[ A.]}, \bibinfo{author}{Hall\xfnm[ P.]},
  \bibinfo{author}{Prvan\xfnm[ T.]}.
\bibinfo{year}{1998.} \newblock \bibinfo{title}{Bandwidth selection for the smoothing of distribution
  functions}.
\newblock \bibinfo{journal}{{\it Biometrika}}
\bibinfo{volume}{{\bf 85}} : \bibinfo{pages}{799--808}.
\bibitem[{Brockwell and Davis(1991)}]{Brockwell91}
\bibinfo{author}{Brockwell\xfnm[ P.]}, \bibinfo{author}{Davis\xfnm[ R.]}.
\bibinfo{year}{1991}.
\newblock \bibinfo{title}{{\it Time Series: Theory and Methods}}.
\newblock \bibinfo{publisher}{Springer Verlag}: \bibinfo{address}{New York}.
\bibitem[{Castellano~M\'endez et~al.(2009)Castellano~M\'endez, Franco,
  Cartelle, Febrero-Bande and Roca}]{castellano2009}
\bibinfo{author}{Castellano~M\'endez\xfnm[ M.]}, \bibinfo{author}{Franco\xfnm[
  A.]}, \bibinfo{author}{Cartelle\xfnm[ D.]},
  \bibinfo{author}{Febrero-Bande\xfnm[ M.]}, \bibinfo{author}{Roca\xfnm[ E.]}.
\bibinfo{year}{2009.} \newblock \bibinfo{title}{Identification of 
$NO_x$ 
and ozone episodes
  and estimation of ozone by statistical analysis}.
\newblock \bibinfo{journal}{{\it Water, Air, \& Soil Pollution}}
\bibinfo{volume}{{\bf 198}} : \bibinfo{pages}{95--110}.
\bibitem[{Castellano-M\'endez et~al.(2004)Castellano-M\'endez,
  Gonz\'alez-Manteiga, Febrero-Bande, Prada-S\'anchez and
  Lozano-Calder\'on}]{castellano04}
\bibinfo{author}{Castellano-M\'endez\xfnm[ M.]},
  \bibinfo{author}{Gonz\'alez-Manteiga\xfnm[ W.]},
  \bibinfo{author}{Febrero-Bande\xfnm[ M.]},
  \bibinfo{author}{Prada-S\'anchez\xfnm[ J.M.]},
  \bibinfo{author}{Lozano-Calder\'on\xfnm[ R.]}.
 \bibinfo{year}{2004.} \newblock \bibinfo{title}{Modelling of the monthly and daily behaviour of the
  runoff of the {X}allas river using {B}ox--{J}enkins and neural networks
  methods}.
\newblock \bibinfo{journal}{{\it Journal of Hydrology}}
\bibinfo{volume}{{\bf 296}} : \bibinfo{pages}{38--58}.
\bibitem[{Celebioglu(2006)}]{celebioglu}
\bibinfo{author}{Celebioglu\xfnm[ T.]}.
\bibinfo{year}{2006.} \newblock \bibinfo{title}{Simulation of Hydrodynamics and Sediment Transport
  Patterns in Delaware Bay}.
\newblock Ph.D. thesis; Drexel University.
\bibitem[{Coles(2001)}]{Coles01}
\bibinfo{author}{Coles\xfnm[ S.]}.
\bibinfo{year}{2001}.
\newblock \bibinfo{title}{{\it An Introduction to Statistical Modeling of Extreme
  Values}}.
\newblock \bibinfo{publisher}{Springer Verlag}: \bibinfo{address}{London}.
\bibitem[{Fan and Gijbels(1996)}]{fan96}
\bibinfo{author}{Fan\xfnm[ J.]}, \bibinfo{author}{Gijbels\xfnm[ I.]}.
\bibinfo{year}{1996}.
\newblock \bibinfo{title}{{\it Local Polynomial Modelling and its Applications}}.
\newblock \bibinfo{publisher}{Chapman \& Hall}: \bibinfo{address}{London}.
\bibitem[{Febrero-Bande and Gonz\'alez-Manteiga(2011)}]{febrero11}
\bibinfo{author}{Febrero-Bande\xfnm[ M.]},
  \bibinfo{author}{Gonz\'alez-Manteiga\xfnm[ W.]}.
  \bibinfo{year}{2011}.
\newblock \bibinfo{title}{Generalized additive models for functional data}.
\newblock In: 
  \bibinfo{booktitle}{{\it Recent Advances in Functional Data Analyisis and Related
  Topics}}, \bibinfo{editor}{Ferraty\xfnm[ F.]} (ed). \bibinfo{publisher}{Physica-Verlag};
 91--96.
\bibitem[{Fern\'andez De~Castro et~al.(2005)Fern\'andez De~Castro, Guillas and
  Gonz\'alez-Manteiga}]{fernandezdecastro2005}
\bibinfo{author}{Fern\'andez De~Castro\xfnm[ B.]},
  \bibinfo{author}{Guillas\xfnm[ S.]},
  \bibinfo{author}{Gonz\'alez-Manteiga\xfnm[ W.]}.
\bibinfo{year}{2005.} \newblock \bibinfo{title}{Functional samples and bootstrap for predicting
  sulfur dioxide levels}.
\newblock \bibinfo{journal}{{\it Technometrics}}
\bibinfo{volume}{{\bf 47}} : 212--222.
\bibitem[{Ferraty et~al.(2005)Ferraty, Rabhi and Vieu}]{sankhya2005}
\bibinfo{author}{Ferraty\xfnm[ F.]}, \bibinfo{author}{Rabhi\xfnm[ A.]},
  \bibinfo{author}{Vieu\xfnm[ P.]}.
\bibinfo{year}{2005.} \newblock \bibinfo{title}{Conditional quantiles for dependent functional data
  with application to the climate {El Nino Phenomenon}}.
\newblock \bibinfo{journal}{{\it Sankhya}}
\bibinfo{volume}{{\bf 67}} : 378--398.
\bibitem[{Ferraty and Vieu(2006)}]{ferratyvieulibro}
\bibinfo{author}{Ferraty\xfnm[ F.]}, \bibinfo{author}{Vieu\xfnm[ P.]}.
\bibinfo{year}{2006}.
\newblock \bibinfo{title}{{\it Nonparametric Functional Data Analysis}}.
\newblock \bibinfo{publisher}{Springer-Verlag}: \bibinfo{address}{New York}.
\bibitem[{Fisher and Tippett(1928)}]{Fisher28}
\bibinfo{author}{Fisher\xfnm[ R.A.]}, \bibinfo{author}{Tippett\xfnm[ L.H.C.]}.
\bibinfo{year}{1928.}
\newblock \bibinfo{title}{Limiting forms of the frequency distributions of the
  largest or smallest member of a sample}.
\newblock \bibinfo{journal}{{\it Proceedings of the Cambridge Philosophical Society}}
\bibinfo{volume}{{\bf 24}} : 180--190.
\bibitem[{Gardes et~al.(2010)Gardes, Girard and Lekina}]{gardes}
\bibinfo{author}{Gardes\xfnm[ L.]}, \bibinfo{author}{Girard\xfnm[ S.]},
  \bibinfo{author}{Lekina\xfnm[ A.]}.
\bibinfo{year}{2010.} \newblock \bibinfo{title}{Functional nonparametric estimation of conditional
  extreme quantiles}.
\newblock \bibinfo{journal}{{\it Journal of Multivariate Analysis}}
\bibinfo{volume}{{\bf 101}} : 419--433.
\bibitem[{Gasser and M\"uller(1984)}]{gasser}
\bibinfo{author}{Gasser\xfnm[ T.]}, \bibinfo{author}{M\"uller\xfnm[ H.G.]}.
\bibinfo{year}{1984.} \newblock \bibinfo{title}{Estimating regression functions and their derivatives
  by the kernel method}.
\newblock \bibinfo{journal}{{\it Scandinavian Journal of Statistics}}
\bibinfo{volume}{{\bf 11}} : 171--185.
\bibitem[{Guo et~al.(1996)Guo, Kachroo and Mngodo}]{Guo96}
\bibinfo{author}{Guo\xfnm[ S.L.]}, \bibinfo{author}{Kachroo\xfnm[ R.K.]},
  \bibinfo{author}{Mngodo\xfnm[ R.J.]}.
\bibinfo{year}{1996. }\newblock \bibinfo{title}{Nonparametric kernel estimation of low flow
  quantiles}.
\newblock \bibinfo{journal}{{\it Journal of Hydrology}}
\bibinfo{volume}{{\bf 185}} : 335--348.
\bibitem[{Hall et~al.(2001)Hall, Poskitt and Presnell}]{hallposkitt2001}
\bibinfo{author}{Hall\xfnm[ P.]}, \bibinfo{author}{Poskitt\xfnm[ D.S.]},
  \bibinfo{author}{Presnell\xfnm[ B.]}.
\bibinfo{year}{2001.} \newblock \bibinfo{title}{A functional data-analytic approach to signal
  discrimination}.
\newblock \bibinfo{journal}{{\it Technometrics}}
\bibinfo{volume}{{\bf 43}} : \bibinfo{pages}{1--9}.
\bibitem[{Hall et~al.(1999)Hall, Rodney and Yao}]{hallwolffyao}
\bibinfo{author}{Hall\xfnm[ P.]}, \bibinfo{author}{Rodney\xfnm[ C.L.]},
  \bibinfo{author}{Yao\xfnm[ Q.]}.
\bibinfo{year}{1999.} \newblock \bibinfo{title}{Methods for estimating a conditional distribution
  function}.
\newblock \bibinfo{journal}{{\it Journal of the American Statistical Association}}
\bibinfo{volume}{{\bf 94}} : 154--163.
\bibitem[{Katz et~al.(2002)Katz, Parlange and Naveau}]{Kat02}
\bibinfo{author}{Katz\xfnm[ R.W.]}, \bibinfo{author}{Parlange\xfnm[ M.B.]},
  \bibinfo{author}{Naveau\xfnm[ P.]}.
\bibinfo{year}{2002.} \newblock \bibinfo{title}{Statistics of extremes in hydrology}.
\newblock \bibinfo{journal}{{\it Advances in Water Resources}}
\bibinfo{volume}{{\bf 25}} : 1287--1304.
\bibitem[{Keskin et~al.(2006)Keskin, Taylan and Terzi}]{Keskin06}
\bibinfo{author}{Keskin\xfnm[ M.E.]}, \bibinfo{author}{Taylan\xfnm[ D.]},
  \bibinfo{author}{Terzi\xfnm[ O.]}.
\bibinfo{year}{2006.} \newblock \bibinfo{title}{Adaptive neural-based fuzzy inference system
  ({ANFIS}) approach for modelling hydrological time series}.
\newblock \bibinfo{journal}{{\it Hydrological Sciences Journal}}
 \bibinfo{volume}{{\bf 51}} : 588--598.
\bibitem[{Kim and Heo(2002)}]{Kim02}
\bibinfo{author}{Kim\xfnm[ K.]}, \bibinfo{author}{Heo\xfnm[ J.]}.
\bibinfo{year}{2002.} \newblock \bibinfo{title}{Comparative study of flood quantiles estimation by
  nonparametric models}.
\newblock \bibinfo{journal}{{\it Journal of Hydrology}}
\bibinfo{volume}{{\bf 260}} : 176--193.
\bibitem[{Koutsoyiannis(2000)}]{Kout00}
\bibinfo{author}{Koutsoyiannis\xfnm[ D.]}.
\bibinfo{year}{2000.} \newblock \bibinfo{title}{A generalized mathematical framework for
stochastic simulation and forecast of hydrologic time series}.
\newblock \bibinfo{journal}{{\it Water Resources
Research}}
\bibinfo{volume}{{\bf 36}} : 1519--1533.
\bibitem[{Lall et~al.(1993)Lall, Moon and Bosworth}]{Lall93}
\bibinfo{author}{Lall\xfnm[ U.]}, \bibinfo{author}{Moon\xfnm[ Y.I.]},
  \bibinfo{author}{Bosworth\xfnm[ K.]}.
\bibinfo{year}{1993.} \newblock \bibinfo{title}{Kernel flood frequency estimators: {B}andwidth
  selection and kernel choice}.
\newblock \bibinfo{journal}{{\it Water Resources Research}}
\bibinfo{volume}{{\bf 29}} : 1003--1015.
\bibitem[{Lall et~al.(1996)Lall, Sangoyomi and Abarbanel}]{Lall96}
\bibinfo{author}{Lall\xfnm[ U.]}, \bibinfo{author}{Sangoyomi\xfnm[ T.]},
  \bibinfo{author}{Abarbanel\xfnm[ H.D.I.]}.
\bibinfo{year}{1996.} \newblock \bibinfo{title}{Nonlinear dynamics of the {G}reat {S}alt {L}ake:
  {N}onparametric short term forecasting}.
\newblock \bibinfo{journal}{{\it Water Resources Research}}
\bibinfo{volume}{{\bf 32}} : 975--985.
\bibitem[{Lall and Sharma(1996)}]{lall_boot}
\bibinfo{author}{Lall\xfnm[ U.]}, \bibinfo{author}{Sharma\xfnm[ A.]}.
\bibinfo{year}{1996.} \newblock \bibinfo{title}{A nearest neighbor bootstrap for resampling
  hydrologic time series}.
\newblock \bibinfo{journal}{{\it Water Resources Research}}
\bibinfo{volume}{{\bf 32}} : 679--693.
\bibitem[{Montanari et~al.(1997)Montanari, Rosso and Taqqu}]{Montanari97}
\bibinfo{author}{Montanari\xfnm[ A.]}, \bibinfo{author}{Rosso\xfnm[ R.]},
  \bibinfo{author}{Taqqu\xfnm[ M.S.]}.
\bibinfo{year}{1997.} \newblock \bibinfo{title}{Fractionally differenced {ARIMA} models applied to
  hydrologic time series: {I}dentification, estimation, and simulation}.
\newblock \bibinfo{journal}{{\it Water Resources Research}}
\bibinfo{volume}{{\bf 33}} : 1035--1044.
\bibitem[{Moon and Lall(1994)}]{MoonLall94}
\bibinfo{author}{Moon\xfnm[ Y.]}, \bibinfo{author}{Lall\xfnm[ U.]}.
\bibinfo{year}{1994.} \newblock \bibinfo{title}{Kernel quantile function estimator for flood
  frequency analysis}.
\newblock \bibinfo{journal}{{\it Water Resources Research}}
\bibinfo{volume}{{\bf 30}} : 3095--3103.
\bibitem[{Parzen(1962)}]{Parzen62}
\bibinfo{author}{Parzen\xfnm[ E.]}.
\bibinfo{year}{1962.} \newblock \bibinfo{title}{On estimation of a probability density function and
  mode}.
\newblock \bibinfo{journal}{{\it Annals of Mathematical Statistics}}
\bibinfo{volume}{{\bf 32}} : 1065--1076.
\bibitem[{Prairie et~al.(2005)Prairie, Rajagopalan, Fulp and
  Zagona}]{Prairie05}
\bibinfo{author}{Prairie\xfnm[ J.R.]}, \bibinfo{author}{Rajagopalan\xfnm[ B.]},
  \bibinfo{author}{Fulp\xfnm[ T.J.]}, \bibinfo{author}{Zagona\xfnm[ E.A.]}.
\bibinfo{year}{2005.} \newblock \bibinfo{title}{Statistical nonparametric model for natural salt
  estimation}.
\newblock \bibinfo{journal}{{\it Journal of Environmental Engineering}}
\bibinfo{volume}{{\bf 131}} : 130--138.
\bibitem[{Quintela-Del-R\'{\i}o(2008)}]{quintela2008}
\bibinfo{author}{Quintela-Del-R\'{\i}o\xfnm[ A.]}.
\bibinfo{year}{2008.} \newblock \bibinfo{title}{Hazard function given a functional variable:
  nonparametric estimation under strong mixing conditions}.
\newblock \bibinfo{journal}{{\it Journal of Nonparametric Statistics}}
\bibinfo{volume}{{\bf 20}} : 413--430.
\bibitem[{Quintela-Del-R\'{\i}o(2011)}]{Quintela2011}
\bibinfo{author}{Quintela-Del-R\'{\i}o\xfnm[ A.]}.
\bibinfo{year}{2011.} \newblock \bibinfo{title}{On bandwidth selection for nonparametric estimation
  in flood frequency analysis}.
\newblock \bibinfo{journal}{{\it Hydrological Processes}}
\bibinfo{volume}{{\bf 25}} : 671--678.
\bibitem[{{R Development Core Team}(2015)}]{Rsoft}
\bibinfo{author}{{R Development Core Team}\xfnm[]}.
\bibinfo{year}{2015}. \newblock \bibinfo{title}{{\it R: A Language and Environment for Statistical
  Computing}}.
\newblock \bibinfo{address}{Vienna, Austria}, 
\newblock \bibinfo{note}{{h}ttp://www.R-project.org}.
\bibitem[{Ramsay and Silverman(2005)}]{ramsaysilvermanlibro}
\bibinfo{author}{Ramsay\xfnm[ J.O.]}, \bibinfo{author}{Silverman\xfnm[ B.W.]}.
\bibinfo{year}{2005}.
\newblock \bibinfo{title}{{\it Functional Data Analysis}}.
\newblock \bibinfo{publisher}{Springer-Verlag}: \bibinfo{address}{New York}.
\bibitem[{Ruppert et~al.(2003)Ruppert, Wand and Carroll}]{Ruppert2003}
\bibinfo{author}{Ruppert\xfnm[ D.]}, \bibinfo{author}{Wand\xfnm[ M.P.]},
  \bibinfo{author}{Carroll\xfnm[ R.J.]}.
  \bibinfo{year}{2003}.
\newblock \bibinfo{title}{{\it Semiparametric Regression}}.
\newblock  \bibinfo{publisher}{Cambridge
  University Press}: \bibinfo{address}{Cambridge, UK.}
\bibitem[{Saf(2009)}]{Saf09}
\bibinfo{author}{Saf\xfnm[ B.]}.
\bibinfo{year}{2009}
\newblock \bibinfo{title}{Regional flood frequency analysis using {L}-moments
  for the {W}est {M}editerranean {R}egion of {T}urkey}.
\newblock \bibinfo{journal}{{\it Water Resources Management}}
\bibinfo{volume}{{\bf 23}} : 531--551.
\bibitem[{Senior and Koerkle(2003)}]{usgs03}
\bibinfo{author}{Senior\xfnm[ L.A.]}, \bibinfo{author}{Koerkle\xfnm[ E.H.]}.
\bibinfo{year}{2003}.
\newblock \bibinfo{title}{Simulation of streamflow and water quality in the
  Christina River subbasin and overview of simulations in other subbasins of
  the Christina River basin, Pennsylvania, Maryland, and Delaware, 1994-98}.
\newblock Number \bibinfo{number}{03-4193} in \bibinfo{series}{{it Water-resources
  investigations report}}.
  \bibinfo{publisher}{U.S. Department of the Interior, U.S. Geological Survey}: \bibinfo{address}{New Cumberland, Pennsylvania}.
\bibitem[{Serinaldi(2009)}]{Serinaldi09}
\bibinfo{author}{Serinaldi\xfnm[ F.]}.
 \bibinfo{year}{2009.} \newblock \bibinfo{title}{Assessing the applicability of fractional order
  statistics for computing confidence intervals for extreme quantiles}.
\newblock \bibinfo{journal}{{\it Journal of Hydrology}}
\bibinfo{volume}{{\bf 376}} : 528--541.
\bibitem[{Sharma et~al.(1997)Sharma, Tarboton and Lall}]{Sharma97}
\bibinfo{author}{Sharma\xfnm[ A.]}, \bibinfo{author}{Tarboton\xfnm[ D.G.]},
  \bibinfo{author}{Lall\xfnm[ U.]}.
\bibinfo{year}{1997.} \newblock \bibinfo{title}{Streamflow simulation: A nonparametric approach}.
\newblock \bibinfo{journal}{{\it Water Resources Research}}
\bibinfo{volume}{{\bf 33}} : \bibinfo{pages}{291--308}.
\bibitem[{Shumway and Stoffer(2011)}]{Shumway}
\bibinfo{author}{Shumway\xfnm[ R.H.]}, \bibinfo{author}{Stoffer\xfnm[ D.S.]}.
\bibinfo{year}{2011}.
\newblock \bibinfo{title}{{\it Time Series Analysis and its Applications}}.
\newblock \bibinfo{publisher}{Springer}: \bibinfo{address}{New York}.
\bibitem[{Singh et~al.(2005)Singh, Wang and Zhang}]{Singh05}
\bibinfo{author}{Singh\xfnm[ V.P.]}, \bibinfo{author}{Wang\xfnm[ S.X.]},
  \bibinfo{author}{Zhang\xfnm[ L.]}.
\bibinfo{year}{2005.} \newblock \bibinfo{title}{Frequency analysis of nonidentically distributed
  hydrologic flood data}.
\newblock \bibinfo{journal}{{\it Journal of Hydrology}}
\bibinfo{volume}{{\bf 307}} : 175--195.
\bibitem[{Smith(1989)}]{Smith89}
\bibinfo{author}{Smith\xfnm[ R.]}.
\bibinfo{year}{1989} \newblock \bibinfo{title}{Extreme value analysis of environmental time series:
  an application to trend detection in ground-level ozone}.
\newblock \bibinfo{journal}{{\it Statistical Science}}
 \bibinfo{volume}{{\bf 4}} : 367--393.
\bibitem[{Tamea et~al.(2005)Tamea, Laio and Ridolfi}]{Tamea05}
\bibinfo{author}{Tamea\xfnm[ S.]}, \bibinfo{author}{Laio\xfnm[ F.]},
  \bibinfo{author}{Ridolfi\xfnm[ L.]}.
\bibinfo{year}{2005.} \newblock \bibinfo{title}{Probabilistic nonlinear prediction of river flows}.
\newblock \bibinfo{journal}{{\it Water Resources Research}}
\bibinfo{volume}{{\bf 41}} : \bibinfo{pages}{W09421--}.
\bibitem[{Toth et~al.(2000)Toth, Brath and Montanari}]{Toth00}
\bibinfo{author}{Toth\xfnm[ E.]}, \bibinfo{author}{Brath\xfnm[ A.]},
  \bibinfo{author}{Montanari\xfnm[ A.]}.
\bibinfo{year}{2000.} \newblock \bibinfo{title}{Comparison of short-term rainfall prediction models
  for real-time flood forecasting}.
\newblock \bibinfo{journal}{{\it Journal of Hydrology}}
\bibinfo{volume}{{\bf 239}} : 132--147.
\bibitem[{Wand and Jones(1995)}]{wan95}
\bibinfo{author}{Wand\xfnm[ M.P.]}, \bibinfo{author}{Jones\xfnm[ M.C.]}.
 \bibinfo{year}{1995}.
\newblock \bibinfo{title}{{\it Kernel Smoothing}}.
\newblock \bibinfo{publisher}{Chapman \& Hall}: \bibinfo{address}{London}.
\bibitem[{Wang et~al.(2009)Wang, Chau, Cheng and Qiu}]{Wang09}
\bibinfo{author}{Wang\xfnm[ W.C.]}, \bibinfo{author}{Chau\xfnm[ K.W.]},
  \bibinfo{author}{Cheng\xfnm[ C.T.]}, \bibinfo{author}{Qiu\xfnm[ L.]}.
\bibinfo{year}{2009.} \newblock \bibinfo{title}{A comparison of performance of several artificial
  intelligence methods for forecasting monthly discharge time series}.
\newblock \bibinfo{journal}{{\it Journal of Hydrology}}
\bibinfo{volume}{{\bf 374}} : \bibinfo{pages}{294--306}.
\bibitem[{Wasserman(2005)}]{Wasserman05}
\bibinfo{author}{Wasserman\xfnm[ L.]}.
 \bibinfo{year}{2005}.
\newblock \bibinfo{title}{{\it All of Nonparametric Statistics}}.
\newblock \bibinfo{publisher}{Springer-Verlag}: \bibinfo{address}{New York}.
\bibitem[{Wu et~al.(2009)Wu, Chau and Li}]{Wu09}
\bibinfo{author}{Wu\xfnm[ C.L.]}, \bibinfo{author}{Chau\xfnm[ K.W.]},
  \bibinfo{author}{Li\xfnm[ Y.S.]}.
\bibinfo{year}{2009.} \newblock \bibinfo{title}{Predicting monthly stream flow using data-driven
  models coupled with data preprocessing techniques}.
\newblock \bibinfo{journal}{{\it Water Resources Research}}
\bibinfo{volume}{{\bf 45}} : \bibinfo{pages}{W08432--}.
\bibitem[{Youndj\'e and Vieu(2006)}]{Youndj06}
\bibinfo{author}{Youndj\'e\xfnm[ E.]}, \bibinfo{author}{Vieu\xfnm[ P.]}.
 \bibinfo{year}{2006.} \newblock \bibinfo{title}{A note on quantile estimation for long dependent
  stochastic processes}.
\newblock \bibinfo{journal}{{\it Statistics \& Probability Letters}}
\bibinfo{volume}{{\bf 76}} : 109--116.

\end{thebibliography}

\end{document}